\documentclass[aps,
prd,
twocolumn,
superscriptaddress,
floatfix
]{revtex4-1}
\usepackage[dvipsnames]{xcolor}
\definecolor{articleColor}{cmyk}{0.3 , 0.3  , 0   , 0.09}  
\usepackage{natbib}
\usepackage{ulem}

\usepackage{amsmath}
\usepackage{amsthm, amsfonts, amssymb}

\usepackage{array}
\usepackage{times}
\usepackage{latexsym}
\usepackage{hyperref}
\hypersetup{backref,  
colorlinks=true,
linkcolor=red,
filecolor=red,
citecolor=blue}
\usepackage{epsfig}
\usepackage{bm}

\newcommand{\D}{\mathrm{d}}

\newcommand{\tr}{\mathrm{tr}}

\newcommand{\Olap}{\mathcal{O}}

\newcommand{\dred}{\textcolor{red!40!black}}

\definecolor{deepdandelion}{RGB}{215,160,65}

\makeatletter
\newcommand{\etal}{\textit{et~al}\@ifnextchar{\relax}{.\relax}{\ifx\@let@token.\else\ifx\@let@token~.\else.\@\xspace\fi\fi}}
\makeatother

\def\l({\left(}
\def\r){\right)}

\newcommand{\CITA}{\affiliation{Canadian Institute for Theoretical
    Astrophysics,
 University of Toronto,
    Toronto, ON M5S 3H8, Canada}} %

\newcommand{\SYR}{\affiliation{Department of Physics, Syracuse University, Syracuse, NY 13244, USA}}
\newcommand{\CIT}{\affiliation{Theoretical Astrophysics 350-17, California Institute of Technology, Pasadena, California 91125, USA}}

\newcommand{\CalState}{\affiliation{Gravitational Wave Physics and Astronomy Center, California State University Fullerton, Fullerton, California 92834, USA}}

\begin{document}

 \author{Prayush Kumar}\thanks{prayush.kumar@ligo.org}\SYR\CITA
 \author{Kevin Barkett}\CIT
 \author{Swetha Bhagwat}\SYR
 \author{Nousha Afshari}\CalState

 \author{Duncan A. Brown}\SYR
 \author{Geoffrey Lovelace}\CalState
 \author{Mark A.~Scheel} \CIT 
 \author{B\'{e}la Szil\'{a}gyi} \CIT 

\date{\today}
\title{Accuracy and precision of gravitational-wave models of inspiraling
neutron star -- black hole binaries with spin: comparison with numerical relativity in the low-frequency regime}
\begin{abstract}
Coalescing binaries of neutron stars and black holes are one of the 
most important sources of gravitational waves for the upcoming network of 
ground based detectors. Detection and extraction of astrophysical information 
from gravitational-wave signals requires accurate waveform models. The Effective-One-Body and 
other phenomenological models interpolate between analytic results and
numerical relativity simulations, that typically span $\mathcal{O}(10)$ orbits
before coalescence. In this paper we study the faithfulness of these models for
neutron star - black hole binaries. We investigate their accuracy using new NR 
simulations that span $36-88$ orbits, with mass-ratios $q$ and black hole spins 
$\chi_\mathrm{BH}$ of $(q,\chi_{BH}) = (7, \pm 0.4), (7, \pm 0.6)$, and $(5, -0.9)$.
We find that:
(i) the recently published SEOBNRv1 and SEOBNRv2 models of the Effective-One-Body
family disagree with each other (mismatches of a few percent) for black hole spins
$\chi_\mathrm{BH}\geq 0.5$ 
or $\chi_\mathrm{BH}\leq -0.3$, with waveform mismatch accumulating during 
\textit{early} inspiral;
(ii) comparison with numerical waveforms indicate that this disagreement is due to 
phasing errors of SEOBNRv1, with SEOBNRv2 in good agreement with all of our 
simulations;
(iii) Phenomenological waveforms agree with SEOBNRv2  only for 
comparable-mass low-spin binaries, with overlaps below $0.7$ elsewhere in the
neutron star - black hole binary parameter space;
(iv) comparison with numerical waveforms shows that most of this model's dephasing 
accumulates near the frequency interval where it switches to a phenomenological phasing
prescription; and finally
(v) both SEOBNR and post-Newtonian models are effectual for neutron star - 
black hole systems, but post-Newtonian waveforms will give a significant bias in
parameter recovery. 
Our results suggest that future gravitational-wave detection searches and parameter estimation efforts
would benefit from using SEOBNRv2 waveform templates when focused on neutron star - black hole
systems with $q\lesssim 7$ and $\chi_\mathrm{BH} \approx [-0.9, +0.6]$.  For larger
black hole spins and/or binary mass-ratios, we recommend the models be further
investigated as NR simulations in that region of the parameter space become available.
\end{abstract}
\maketitle
\section{Introduction}


The Advanced Laser Interferometer Gravitational-wave Observatory
(aLIGO)~\cite{0264-9381-32-7-074001,Harry:2010zz} is currently being
commissioned and will begin observation in 2015, reaching its design 
sensitivity by $2018-19$. The Virgo gravitational-wave observatory~\cite{aVIRGO}
will begin operation in $2016$. With improved 
sensitivity, these detectors will access a thousand times as much volume as their first generation counterparts.
In addition, the KAGRA detector is currently under construction in 
Japan~\cite{Somiya:2011np}, and a plan to build an advanced LIGO detector 
in India is under consideration.
Compact binaries are the most promising sources of gravitational waves (GWs) 
for aLIGO. Binary systems containing stellar-mass black holes (BH) and/or 
neutron stars (NS) inspiral and merge because of their GW emission. 
The GW waves emitted with frequencies above $\sim 10$~Hz will be in the sensitive
band of aLIGO and Virgo.

In this paper, we focus on neutron star -- black hole
(NSBH) binaries. Based
on our current understanding of the astrophysical NS and BH population, stellar
binary evolution, and on population synthesis studies, we expect aLIGO to observe
$0.2-300$ NSBH binary mergers a year~\cite{LSCCBCRates2010}. 
GW observations of NSBH binaries have significant scientific potential, beyond
the initial discovery of a new class of astrophysical systems. GWs emitted by 
coalescing NSBH binaries carry signatures of strong-field gravitational 
dynamics. Unlike binary neutron stars, GWs from NSBH binaries will contain 
the signatures of the interaction of 
BH spins~\cite{2011ApJ...742...85G,2012MNRAS.424..217F,Gou:2013dna,
2009ApJ...697..900M,McClintock:2006xd,Miller:2009cw} 
with the orbital motion. Significant efforts are underway to access this 
information using the aLIGO and Virgo detector network to test general 
relativity in the strong gravity regime~\cite{Agathos:2013upa,DelPozzo:2014cla}.
The observation and characterization of a population of NSBH sources will 
also shed light on stellar evolution and compact-binary formation 
mechanisms: e.g., a gap in the mass distribution of NSs and 
BHs could shed light on the mechanism of supernova 
explosions~\cite{IlyaCBCClassification,Dominik:2013tma,Fryer:2011cx}.
An unambiguous detection of  GWs from a NSBH system accompanied by 
electromagnetic observations could provide information about the internal
structure of NSs~\cite{Pannarale:2014rea} and could provide strong evidence
linking compact binary mergers and short Gamma-ray bursts
(SGRBs)~\cite{1984SvAL...10..177B,Paczynski:1986px,Eichler:1989ve,Foucart:2012vn}.
However, unlocking the full scientific potential of GWs emitted by NSBH
coalescences requires both detecting as many of such signals as possible and 
accurately characterizing them to understand the properties of their source
binaries.


Detection searches are based on the matched-filtering technique~\cite{1057571},
using modeled waveforms as filter templates. Searches for compact binaries
with initial LIGO and Virgo detectors used non-spinning template waveforms~\cite{VanDenBroeck:2009gd,Colaboration:2011nz,Abadie:2010yb,
Abbott:2009qj,Abbott:2009tt} (with the exception of~\cite{Abbott:2007ai},
\dred{ChrisVDBrock et al}).
While the cataloged astrophysical population of NSs have small spins
(mass-normalized $|\vec{\chi}|\lesssim 0.05$), the spins of
stellar-mass BHs are uncertain, with estimates ranging from low to nearly 
extremal values (i.e., nearly as fast as possible---see, 
e.g.~\cite{McClintockEtAl:2006,Miller:2009cw,Gou:2014una} for examples of 
nearly extremal estimates of BH spins, and see 
Refs.~\cite{McClintock:2013vwa,Reynolds:2013qqa} for recent reviews of 
astrophysical BH spin measurements).
Recent work has shown that including 
non-precessing (that is, aligned) component spins in templates 
used in matched-filtering gravitational-wave searches
will significantly
improve the searches' sensitivity~\cite{Harry:2013tca}. Therefore, 
aLIGO-Virgo searches targeting NSBH binaries plan to use aligned-spin waveform
templates~\cite{Canton:2014ena}.
Because they are central to matched-filtering searches,
it is crucial to have GW models that
accurately capture the NSBH coalescence process. 
Modeling inaccuracy would reduce the signal-to-noise ratio (SNR) recovered by
detection searches, and degrade the range of aLIGO-Virgo observatories.
It would also lead to systematic, but not necessarily controlled, biases in the 
recovered masses and spins of the source.

Studies of the accuracy of contemporary waveform models in the past
have focused on post-Newtonian (PN)~\cite{PNtheoryLivingReviewBlanchet}
and recent Effective-One-Body (in particular, the ``SEOBNRv1''~\cite{Taracchini:2012ig}) models.
It has been shown that PN approximants disagree significantly with
each other and with SEOBNRv1 for aligned-spin NSBH binaries~\cite{Nitz:2013mxa}, despite the inclusion of the highest-known order spin 
contributions to the binary phasing~\cite{Bohe:2013cla,Blanchet:2011zv}. While the 
accuracy of the SEOBNR models is enhanced through calibration against 
high-accuracy Numerical Relativity (NR) merger simulations, most of these 
simulations correspond to comparable mass ratios. Therefore, the extension of 
SEOBNR into NSBH parameter space is not guaranteed to be reliable.

In this paper, we systematically investigate waveform approximants in the
context of NSBH binaries. Unlike past studies, we investigate not just the 
\textit{precision} (mutual agreement of approximants) but also their
\textit{accuracy}, by comparing with long
NR simulations with $q=m_{\mathrm{BH}}/m_{\mathrm{NS}}=\{5,7\}$ and aligned
BH spin $\chi_\mathrm{BH}=S_\mathrm{BH}/M_\mathrm{BH}^2 = \{\pm 0.4, \pm 0.6, -0.9\}$. 
(Note that, except where we specify otherwise, we adopt geometrized 
units with $G=c=1$ in this paper). These simulations are 
described further in Sec.~\ref{s1:NRwaveforms}.
In addition to PN and SEOBNRv1, we compare with the more recent SEOBNRv2 and the 
phenomenological PhenomC~\cite{Santamaria:2010yb} models. 
We use the zero-detuning high power noise curve for Advanced 
LIGO~\cite{aLIGONoiseCurve} with a $15$~Hz lower frequency cutoff in our 
calculations. We allow the BH spin to vary over $[-1,1]$, and its mass to vary 
over $[3M_\odot,14M_\odot]$. The NS mass is fixed at
$m_\mathrm{NS}=1.4M_\odot$ with $\chi_\mathrm{NS}=0$, as is consistent
with the observed astrophysical NS population~\cite{2013ApJ...778...66K,
1993MNRAS.263..403L,Postnov:2014tza}. 
Note that while investigating waveform modeling errors, we ignore NS matter effects
and treat the NS as a low-mass BH. Although matter effects are expected to be
measurable by aLIGO (e.g.~\cite{Read:2013zra,Lackey:2013axa}), they affect the
inspiral phasing starting at $5$PN order. As there are lower-order spin-dependent
vacuum terms in PN phasing that remain unknown, the effect of ignoring 
matter-dependent secular terms will be sub-dominant to other sources of error 
in waveform models.
We also ignore the effect of NS disruption before merger, which is likely
when the mass-ratio $m_\mathrm{BH}/m_\mathrm{NS}$ is small and/or the BH
has relatively high aligned spin~\cite{Foucart:2014nda}.
However, this disruption occurs at fairly high frequencies, i.e. at
$f_\mathrm{GW}\gtrsim 1.2$~kHz~\cite{Foucart:2014nda}, and its effects are
expected to be small due to the significantly reduced sensitivity of aLIGO
at such frequencies~\cite{aLIGOsensitivity}. 
We leave the study of this effect to future work.

First, we study GW model \textit{precision} by comparing the PN time-domain 
TaylorT4, PN frequency-domain TaylorF2, SEOBNRv1 and PhenomC models 
with the most recent SEOBNRv2 model. As SEOBNRv2 has been calibrated to 
$38$ NR simulations, we take it as the fiducial model representing the true 
waveform.
We find that both PN models have overlaps with SEOBNRv2 \textit{below} $0.9$ for mass-ratio 
$q\geq 3$ and/or BH spin $|\chi_\mathrm{BH}|\geq 0.5$.
We also find that PhenomC and SEOBNRv2 have overlaps
below $0.9$ for $q\leq 5$ and/or BH spin $|\chi_\mathrm{BH}|\geq 0.3$, falling as low as $0.6$.
Finally, we also find the overlaps between SEOBNRv1 and SEOBNRv2 fall below $0.9$ 
for NSBH systems with anti-aligned BH spins $\chi_\mathrm{BH}\leq -0.5$.

We further investigate the \textit{accumulation of mismatch} between different models,
as a function of GW frequency. For PN approximants, we find that most of the 
mismatch is accrued during the late-inspiral phase when the PN velocity 
parameter $v/c\gtrsim 0.2$.  This is expected, because PN results are 
perturbative expansions in $v/c$ that break down when $v$ becomes comparable
to $c$ near the time of coalescence.
Between SEOBNRv1 and SEOBNRv2, we find that mismatch is accrued 
during the early-to-late inspiral transition period when $v/c \lesssim 0.26$. 
We find a similar trend between PhenomC and SEOBNRv2.
This demonstrates discrepancies between
NR-calibrated models in the early inspiral phase, despite good agreement
close to merger, where all of the models have been calibrated to NR.

Second, we study the \textit{accuracy} of NSBH waveform models by computing 
their overlaps (or \textit{faithfulness}) against our long NR simulations. 
Our simulations extend down to $v/c\simeq 0.2$, and
are long enough to probe the frequency range
in which SEOBNRv1/PhenomC phase evolutions differ from SEOBNRv2. 
%
We find that SEOBNRv1 has $1-3\%$ mismatches against the aligned-spin
simulations, which rise up to $\sim 5\%$ against the anti-aligned-spin ones.
While most of this mismatch is accumulated during the last few pre-merger orbits for 
\textit{aligned}-spin cases, for \textit{anti-aligned} cases it accumulates over the
$30-50$ inspiral orbits that our simulations span.
On the other hand, we found that SEOBNRv2 has $<1\%$ mismatches with NR, 
for both aligned and anti-aligned simulations. Therefore, we conclude that the 
differences between the two SEOBNR models are because of the phasing errors 
in SEOBNRv1. 
For PhenomC and both PN approximants, we find $\geq 10\%$ mismatches 
against NR, for both aligned and anti-aligned spin simulations. 
%
We therefore conclude that SEOBNRv2 provides the most accurate description
of aligned-spin NS-BH coalescence waveforms, with the caveat that the model
should be analyzed for more extreme component spins.

Third, we investigate the \textit{suitability of different models for detection searches}.
To address this question, we compute the \textit{effectualness} of different models by
allowing the additional degree of freedom of maximizing overlaps between analytic and
numerical waveforms over intrinsic binary parameters.
 We find that both SEOBNRv1 and SEOBNRv2 recover $\geq 99.8\%$ of the optimal SNR.
PhenomC shows low SNR recovery, which drops below $\sim 90\%$ for
anti-aligned BH spins. We therefore recommend against using this model 
in NSBH detection searches. 
Both PN models recover about $98\%$ of the SNR for aligned-spin systems,
and are therefore effectual for aLIGO searches. For anti-aligned systems, both TaylorT4 and TaylorF2 
models recover $\lesssim 96\%$ of the SNR and would likely benefit
from the computation of higher order spin-dependent corrections to PN
dynamics.
%
Therefore we recommend that SEOBNRv2 be preferred in aLIGO NSBH detection 
searches.

Finally, we probe the question of \textit{systematic biases in parameter recovery} 
corresponding to using each approximant to model aLIGO parameter estimation templates.
We find that the accuracy of the chirp mass 
$\left(\mathcal{M}_c=(m_1m_2)^{3/5}(m_1+m_2)^{-1/5}\right)$ recovery 
increases with the number of orbits that are integrated over. All approximants 
recovered $\mathcal{M}_c$ within a few percent.
The spin--mass-ratio degeneracy makes the accurate determination
of mass-ratio and component spins more challenging. We find systematic
biases in mass-ratio to be between a few to tens of percents, increasing
with BH spin, with similar biases in the recovered values of BH spins.
%
Of all the models considered, we find that SEOBNRv2 surpasses others in
faithfulness.
%
Using the accuracy measures proposed in~\cite{WaveformAccuracy2008}, we also 
found SEOBNRv2 to be indistinguishable from true waveforms up to SNRs 
$\approx 8-14\, (16-18)$ for aligned (anti-aligned) BH spins.
We therefore recommend that SEOBNRv2 be used in aLIGO parameter
estimation efforts for aligned-spin NSBH detection candidates, but we 
also recommend that SEOBNR be tested for higher component spins.

Our results are limited by the fact that our NR waveforms only extend down to 
$v/c\simeq 0.21-0.24$ (i.e. $60-80$~Hz for NSBH masses), while aLIGO is 
sensitive
down to $15$~Hz. A sizable fraction ($35-45\%$, depending on BH spin)
of the signal power will be accumulated at frequencies below this range. We
plan to extend these calculations to lower frequencies in future work. 
%
%
%

The remainder of the paper is organized as follows: 
In Sec.~\ref{s1:NRwaveforms}, we describe the NR waveforms presented in this 
paper and discuss their convergence. 
In Sec.~\ref{s1:waveforms}, we describe the waveform models 
studied here. 
In Sec.~\ref{s1:quantifyingerrors}, we describe the measures used to 
quantify waveform discrepancies.
In Sec.~\ref{s1:modelcomparison}, we discuss the faithfulness of different
waveform approximants for different NSBH masses and spins, and also
as a function of the emitted GW frequency.
In Sec.~\ref{s1:numrelcomparison}, we investigate the late-inspiral
accuracy of all approximants using our high accuracy numerical 
simulations.
In Sec.~\ref{s1:parameterbias}, we study the viability of using different
approximants as detection templates, as well as their intrinsic parameter 
biases for aLIGO parameter estimation studies.
In Sec.~\ref{s1:conclusions}, we summarize and discuss our results.

\section{Numerical Relativity simulations}\label{s1:NRwaveforms}

\begin{table*}
\begin{tabular}{| c | c | c | c | c | c | c | c | c |}
\hline
ID & q & $\vec{\chi}_1$ & No. of orbits & Initial $f_{gw}$ (Hz) & Initial $M\omega_{orbital}$ & $\dot{a}$ & $D_0$ & $\epsilon$ \\ \hline
SXS:BBH:0202 & 7 & $(0, 0,  0.6)$ & 62.1 & 75.5 & 0.01309 & $4.3970\times10^{-5}$ & 17.0000 & $9\times10^{-5}$ \\
SXS:BBH:0203 & 7 & $(0, 0,  0.4)$ & 58.5 & 76.0 & 0.01317 & $-9.8403\times10^{-6}$ & 17.0005 & $<1.6\times10^{-4}$ \\
SXS:BBH:0204 & 7 & $(0, 0,  0.4)$ & 88.4 & 60.3 & 0.01045 & $-4.6373\times10^{-6}$ & 20.0000 & $<1.7\times10^{-4}$ \\
SXS:BBH:0205 & 7 & $(0, 0,  -0.4)$ & 44.9 & 76.0 & 0.01318 & $-1.4760\times10^{-5}$ & 17.1036 & $7.0\times10^{-5}$ \\
SXS:BBH:0206 & 7 & $(0, 0,  -0.4)$ & 73.2 & 59.8 & 0.01036 & $-7.8300\times10^{-6}$ & 20.2167 & $<1.6\times10^{-4}$ \\
SXS:BBH:0207 & 7 & $(0, 0,  -0.6)$ & 36.1 & 80.8 & 0.01399 & $7.1708\times10^{-6}$ & 16.4000 & $1.693\times10^{-4}$ \\
SXS:BBH:0208 & 5 & $(0, 0,  -0.9)$ & 49.9 & 80.0 & 0.0104 & $-4.5088\times10^{-5}$ & 20.0778 & $5.074\times10^{-4}$ \\
\hline
\end{tabular}
\caption{Numerical-relativity simulations used in this study (each
  performed using SpEC~\cite{spec}). For each simulation (labeled by ID), the
  table shows the mass ratio $q\equiv m_1/m_2\geq 1$, the spin $\vec{\chi}_1$ of
  the heavier compact object (the lighter object is non-spinning), the number of
  orbits, the initial gravitational-wave frequency $f_{gw}$ when the total mass
  is scaled so that the system mimics a NSBH binary with a NS mass of
  $1.4M_\odot$, the initial dimensionless orbital velocity $M \omega_{orbital}$,
  radial velocity $\dot{a}$, separation $D_0$, and eccentricity $\epsilon$.
  These values listed for $f_{gw}, M\omega_{orbital}$, $\dot{a}$ and $D_0$ are
  for the initial data, before junk radiation. The $\vec{\chi}_1=\pm0.4$
  simulations use CFMS initial data while the rest use SKS initial data.}
\label{table:simlist}
\end{table*} 

\begin{figure*}
\includegraphics[width=7in]{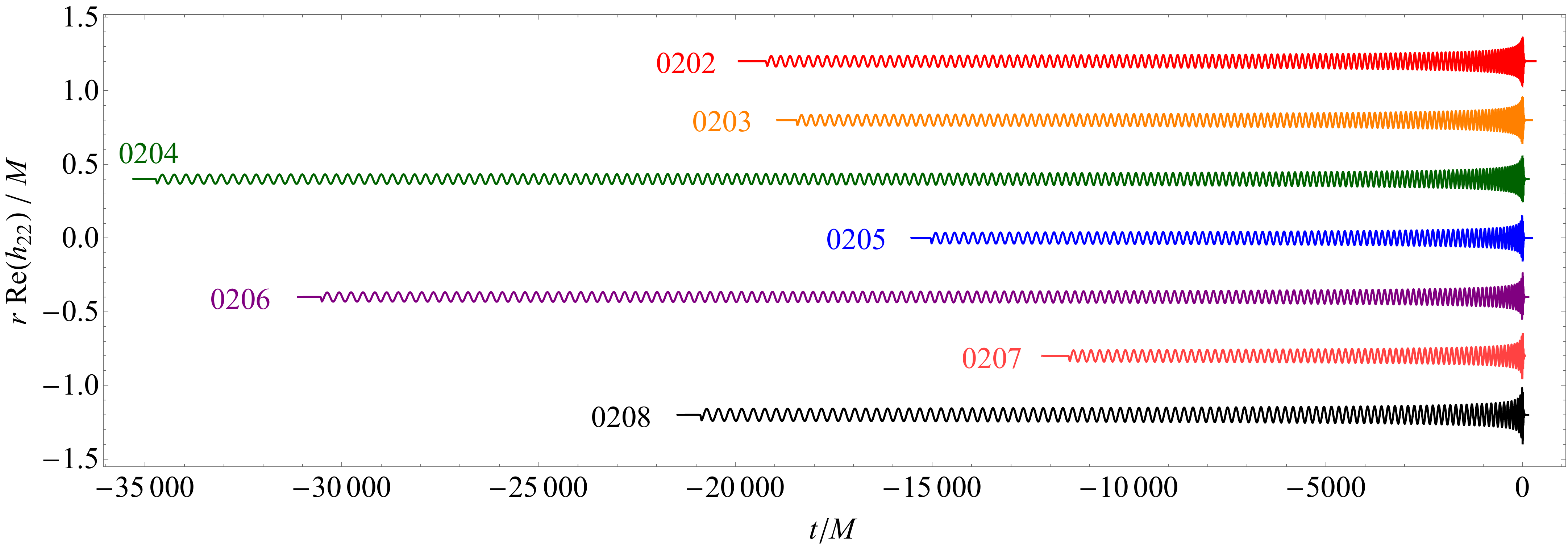}
\caption{The real part of $r h_{22} / M$ of the $\ell=m=2$ mode of the numerical 
waveforms used in this paper, where $M$ is the total mass and $r$ is the 
radial distance from the source to an observer. The waveform labeled by 
$N$ correspond to simulation SXS:BBH:$N$, where $N\in\left\{202,203,204,205,206,207\right\}$.
The waves are shown as 
a function of time $t$. A constant vertical offset is applied to each 
waveform for clarity, and the waves are offset in time so the peak amplitude 
occurs at time $t=0$.
\label{fig:Ccewaveform_plot}
}
\end{figure*}

\begin{figure}
\includegraphics[width=\columnwidth]{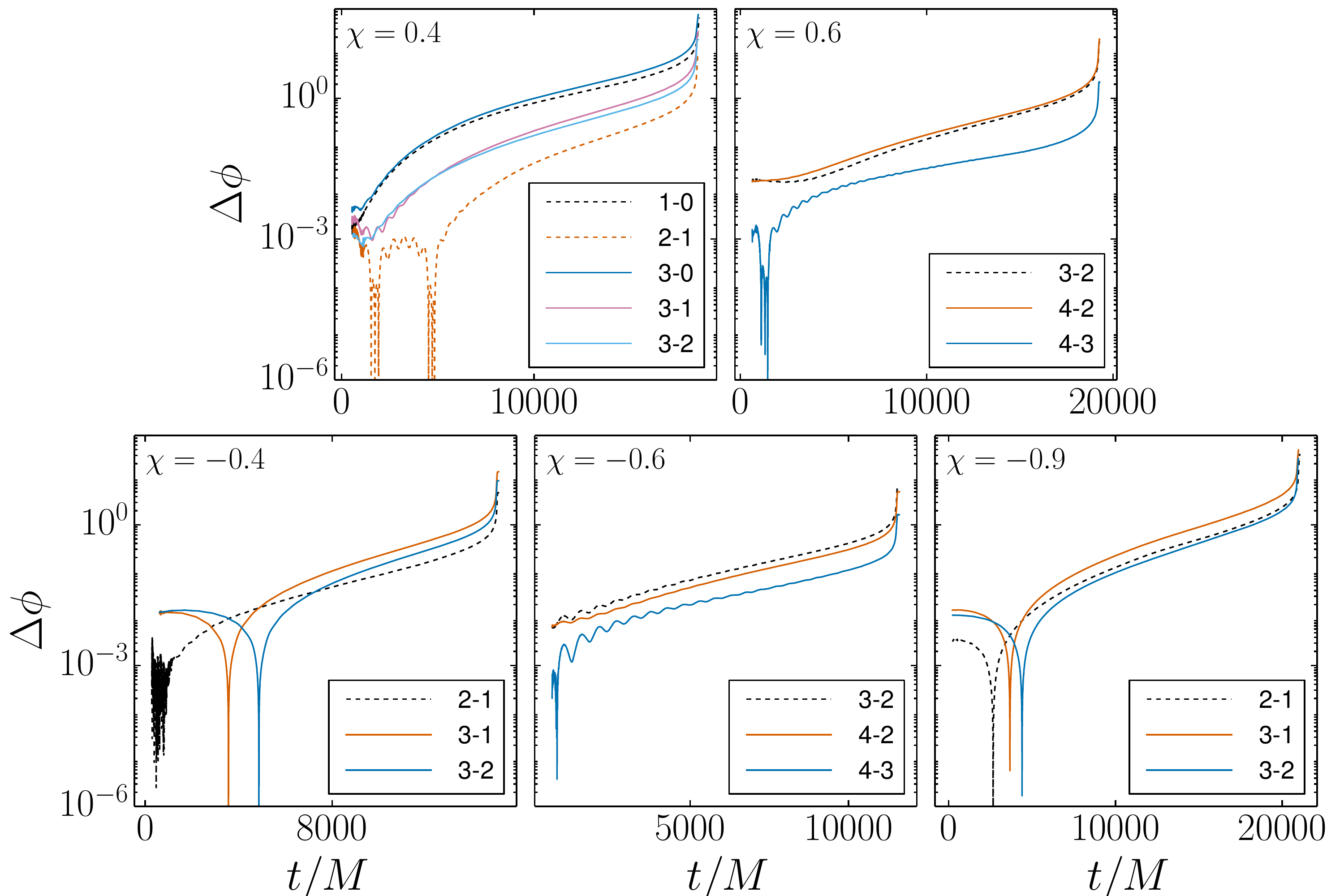}
\caption{Phase differences between different resolutions for the $\ell=2,m=2$
modes of $\Psi_4$, plotted as a function of time. Only the five numerical
simulations with more than 2 resolutions are displayed. The legends indicate
which resolutions are compared, e.g. ``3-2'' compares the phase of $N=3$ versus
$N=2$.
\label{fig:ConvergenceTest}
}
\end{figure}

We construct our NR waveforms using the Spectral Einstein Code (SpEC)
~\cite{spec}. Their parameters are summarized in Table~\ref{table:simlist}.
Of particular note is the length of these simulations; the shortest
waveform presented here has over 36 orbits of inspiral before the merger and the
longest has nearly 90 orbits. Fig.~\ref{fig:Ccewaveform_plot} shows the real
part of the waveform dimenionless strain, $rh_{22}/M$ for each of the
simulations on Table~\ref{table:simlist}. With the exception of the 176 orbit simulation
presented in~\cite{Szilagyi:2015rwa} and a $48.5$ orbit simulation presented
in~\cite{Lousto:2015a} these waveforms are among the longest done to date. The longest
waveform currently in the SXS catalog is only 35.5 orbits ~\cite{sxscatalog}.

To test the accuracy of the simulations we ran each simulation using
different numerical resolutions; we label each resolution by an integer $N$,
where larger $N$ indicates finer resolution. We compute the phase of the
$\ell=2,m=2$ mode of $\Psi_4$ (the second time derivative of the complex strain
$h_{22}(t)$) for different resolutions
Fig.~\ref{fig:ConvergenceTest} shows these phase differences for each pair of resolutions for five
of the seven numerical simulations presented here. (We ran the other two
simulations at fewer than 3 values of $N$, so such
a plot would not be useful for those cases.)

If for all subdomains, i) the number of grid points increased
uniformly with increasing $N$, and ii) at any given time, the
locations of the boundaries of all subdomains were independent of $N$, 
then Fig.~\ref{fig:ConvergenceTest} would represent a classic
convergence test. In that case, the phase differences should decrease with $N$
in a predictable way, according to the convergence order of the
numerical scheme.  The simulations here, however, use a spectral adaptive mesh
refinement (AMR) scheme~\cite{Szilagyi:2014fna}, and the label $N$ determines the error
tolerance used by AMR when it decides whether to change the number of
points in a given subdomain and when it decides whether to split a
single subdomain into many smaller ones or to join several subdomains
into a larger one.  Because AMR makes these decisions independently
for different values of $N$, at any given time it is possible that a
given subdomain has the same number of grid points 
for two values of
$N$, and it is possible that subdomain boundaries for different values
of $N$ do not agree.  Therefore, we do not necessarily
expect strict convergence in Fig.~\ref{fig:ConvergenceTest}.
These issues will be discussed in detail in a separate paper that focuses
on convergence of BBH runs using SpEC.

Nevertheless, for the $\chi= \pm 0.6$ simulations, 
the differences converge well with $N$: differences become successively
smaller with increasing resolution. 
Furthermore, the difference between $N=3$ and $N=2$ is approximately
equal to the difference between $N=4$ and $N=2$, 
indicating that these differences essentially measure the error in $N=2$.
For the $\chi = \pm 0.4$ and $\chi=-0.9$ simulations, the difference
between $N=3$ and $N=2$ is smaller than the difference between $N=3$ and $N=1$,
but the spacing between differences is not uniform, and there are some
anomalously small phase differences, such as between $N=2$ and $N=1$ for
$\chi = 0.4$.

For $\chi= \pm 0.6$, the difference between the two finest resolutions
$N=3$ and $N=4$ is a
good measure of the numerical error in the $N=3$
simulation.  The error in $N=4$ could be similarly measured via the difference
between $N=4$ and $N=5$, but since we do not have
an $N=5$ simulation, we take the difference between
$N=3$ and $N=4$ as an extremely conservative estimate of the error in $N=4$.  
For $\chi = \pm 0.4$ and $\chi=-0.9$, where the convergence
with $N$ is not so apparent, we likewise take the difference between the
two highest resolutions as the error estimate of the highest-resolution
simulation; in these cases the error estimate is likely not
as conservative as for $\chi= \pm 0.6$.

\section{Waveform Approximants}\label{s1:waveforms}

In this paper, we consider three waveform families: post-Newtonian, 
Effective-One-Body, and phenomenological 
models~\cite{Ajith:2009bn,Santamaria:2010yb}.
We briefly summarize them here, pointing the reader to the references for  
more detailed descriptions. 
We consider the $(\ell,m) = (2,\pm 2)$ spin-weighted spherical 
harmonic waveform multipoles, since (i) these are the dominant modes for
non-precessing systems, with the contributions from other modes being
smaller by a few orders of magnitude, and (ii) none of the contemporary
IMR models include the sub-dominant waveform modes. 

\textit{\bf Post-Newtonian:}
The Post-Newtonian (PN) approximation is a perturbative expansion of 
compact binary inspiral dynamics in the limit of slow motions and weak fields.
The orbital energy $E$ of a non-precessing binary and the flux $F$ of 
the gravitational energy emitted as GWs are known to $3.5$PN 
order~\cite{Sathyaprakash:1991mt,Cutler:1994ys,Droz:1999qx,Blanchet:2004ek,
Blanchet:2004bb,Jaranowski:1999qd,Jaranowski:1999ye,Damour:2001bu,Blanchet3PN}.
Combining $E$ and $F$ using the energy balance equation $dE/dt = -F$
yields a system of differential equations; solving these equations 
gives the GW phase and the orbital frequency evolution.
The energy balance equation can be 
re-expanded and solved in
different ways to obtain different approximants that agree to 3.5PN order 
but differ at higher orders. The PN formalism and the corresponding 
equations of motion 
break down before merger as the underlying approximations (slow motions and 
weak fields) break down; therefore, the PN formalism produces only the inspiral
portions of the waveform.
In this paper, we will examine two particular PN approximants, the time-domain
TaylorT4 and frequency-domain TaylorF2. In both, we include the recently 
published spin-orbit tail (3PN) and the next-to-next-to-leading order 
spin-orbit (3.5PN) contributions~\cite{Blanchet:2012sm,Bohe:2012mr}.
We refer the reader to the Appendix of 
Ref.~\cite{Nitz:2013mxa} for a summary of the expressions that describe both 
approximants.

\textit{\bf Effective-One-Body:}
The Effective-One-Body (EOB) approach solves for the dynamics of a compact 
binary system by mapping them to the dynamics of an effective test particle of mass 
$\mu=m_{1}m_{2}/(m_{1}+ m_{2})$ with a spin 
$S^{*}(m_{1}, m_{2}, S_{1}, S_{2})$ in a space-time described by a suitable
deformation of the Kerr metric. 
Specifically, when constructing model waveforms using the EOB approach, 
one chooses the deformation and 
the test-particle spin such that the
geodesic followed by the test particle reproduces the PN-expanded
dynamics of the compact binary system with component masses $m_{1}$
and $m_{2}$ and spins $S_{1}$ and $S_{2}$.  Then, one 
matches the coefficients of the
deformed metric to the PN expansion up to 3PN
order; to further improve accuracy, 
one adds adjustable 4PN and 5PN terms that are
calibrated by forcing agreement with NR waveforms. 
The conservative dynamics of the test particle in the deformed-Kerr spacetime 
are described by the EOB Hamiltonian $H_\mathrm{EOB}$. The expressions for 
$H_\mathrm{EOB}$ for different Spinning-EOB (SEOBNR) models differ at high PN 
orders and can be found in Refs.~\cite{Taracchini:2012ig,Taracchini:2013rva}.
Further, a radiation reaction
term in the equations of motion captures the non-conservative dynamics, i.e., 
%
%
the motion of the binary through inspiral to
merger~\cite{Taracchini:2012ig,Taracchini:2013rva}.
In contrast to PN waveforms, EOB waveforms continue through merger and 
ringdown, with the ringdown waveform constructed as a linear superposition of 
the first eight 
quasi-normal modes (QNMs) of the Kerr BH formed at binary
merger~\cite{2009PhRvD..79l4028B}. Matching the ringdown waveform to the 
inspiral-merger portion determines the coefficients associated with these
QNMs.

We use two SEOBNR models (both available in the 
LIGO Algorithm Library (LAL)~\cite{lal}) in this study: SEOBNRv1 and SEOBNRv2. These models 
differ in their calibration to NR waveforms. SEOBNRv1 models the
inspiral-merger-ringdown (IMR) waveform for binaries with component spins 
$-1\leq\chi_{1,2}\leq0.6$; while SEOBNRv2 can model more extremal component
spins, i.e. $-1\leq\chi\leq0.99$. 
SEOBNRv1 uses 5 non-spinning NR simulations at mass ratios $q=\{1,2,3,4,6\}$
and two equal-mass simulations with $\chi_{1,2}=\pm 0.4$ to 
choose six adjustable parameters; note that NSBH systems are outside the 
domain of calibration of this model. 
SEOBNRv2, in addition, includes an adjustable parameter in the effective
particle spin mapping, one in the Hamiltonian, and one
in the complex phase of $h_{2,2}$. These parameters have been
calibrated against 8 non-spinning and 30 aligned-spin NR waveforms. We refer the
readers to Ref.~\cite{Taracchini:2012ig} and Ref.~\cite{Taracchini:2013rva} 
for comprehensive descriptions of SEOBNRv1 and SEOBNRv2, respectively.

\textit{\bf Phenomenological model:}
PhenomC is a phenomenological model that has a closed form in the frequency 
domain and describes the GW emitted by aligned spin binaries during their 
inspiral-merger-ringdown (IMR) phases~\cite{Santamaria:2010yb}. 
Closed form TaylorF2 expressions capture the early adiabatic inspiral stage 
of binary coalescence, with the frequency-domain waveform 
amplitude and phase expressed as expansions in GW frequency.
PhenomC inspiral phasing includes non-spinning terms up to $3.5$PN order,
spin-orbit and spin-spin coupling terms up to $2.5$PN order, and horizon
absorption terms up to $2.5$PN order~\cite{Alvi:2001mx}. 
%
In the late-inspiral/pre-merger regime, the PN approximation 
is insufficient to model the phase 
evolution accurately; instead, a phenomenologically fitted power series
in 
frequency (i.e., a polynomial in $f^{1/3}$) captures the phase evolution. 
%
%
Calibrating against a set of NR 
waveforms~\cite{Santamaria:2010yb} in the frequency range
$[0.1 f_\mathrm{RD}, f_\mathrm{RD}]$, where $f_\mathrm{RD}$ is the primary
(least damped) GW frequency emitted during the quasi-normal ringing of the
post-merger black hole remnant, determines the free coefficients in the 
resulting \textit{pre-merger} phase prescription
In the ringdown regime, PhenomC models the phase 
as a linear function in GW frequency, 
capturing the effect of the leading quasi-normal mode. 
Similarly, PhenomC constructs the amplitude prescription through piece-wise 
modeling of the pre-merger and ringdown regimes, approximating the 
amplitude by a power-series in 
$f^{1/3}$ pre-merger and by a Lorentzian post-merger. 
%
For a complete description of PhenomC and its calibration, we refer the 
reader to Ref.~\cite{Santamaria:2010yb}. Note that PhenomC
belongs to the unique class of models that are both closed-form in the 
frequency domain, \textit{and} include the late-inspiral, merger, and ringdown
in the waveforms.
These features are especially convenient for real GW searches, which operate in
the frequency domain, filtering observatory data with a
large number $\left(10^5-10^6\right)$ of waveform templates.


\section{Quantifying waveform accuracy}\label{s1:quantifyingerrors}

As a measure of how ``close'' two waveforms $h_1$ and $h_2$ are in the waveform
manifold, we use the maximized overlap (match) $\mathcal{O}$, defined by 
\begin{equation}\label{eq:maxnormolap}
\Olap(h_1,h_2) \equiv \dfrac{\underset{\phi_c,t_c}{\mathrm{max}}\,\l(h_1|h_2(\phi_c,t_c)\r)}{\sqrt{(h_1|h_1)(h_2|h_2)}},
\end{equation}
where the overlap $\l(\cdot|\cdot\r)$ between two waveforms is 
\begin{equation}\label{eq:overlap}
(h_1|h_2) \equiv 
4\int^{f_\mathrm{Ny}}_{f_\mathrm{min}}\dfrac{\tilde{h}_1(f)\tilde{h}_2^*(f)}{S_n(f)}\D f;
\end{equation}
$\phi_c$ and $t_c$ are the phase and time shift differences between $h_1$ and 
$h_2$; $\tilde{h}(f)$ is the Fourier transform of the GW waveform $h$;
$S_n(f)$ is the one-sided power spectral density (PSD) of the detector noise, 
which we assume to be stationary and Gaussian with zero mean; 
$f_\mathrm{min}$ is the lower frequency cutoff for filtering; and 
$f_\mathrm{Ny}$ is the Nyquist frequency corresponding to the waveform 
sampling rate. The normalization of $\mathcal{O}$ takes away the effect of any 
overall amplitude scaling differences between $h_1$ and $h_2$.  
The complimentary measure of the \textit{mismatch} $\mathcal{M}$ 
between the two waveforms is therefore
\begin{equation}\label{eq:mismatch}
\mathcal{M}(h_1,h_2) = 1 - \Olap(h_1,h_2).
\end{equation}

Matched-filtering based searches use a discrete bank of modeled
waveforms as filters. The optimal value of the recovered SNR is
$ \rho_{\mathrm{opt}} = \sqrt{\l(h^\tr|h^\tr\r)}$, where $h^\tr$ is
the actual GW signal in the detector output data.
With a finite bank of filter templates, the recovered SNR is
$\rho\simeq \Olap(h^{\tr},h_b)\,\rho_{\mathrm{opt}}\leq \rho_{\mathrm{opt}}$,
where $h_b$ is the filter template in the bank that has the highest maximized
overlap with the signal $h^{\tr}$; i.e. the recovered SNR is maximized over 
both intrinsic (component mass and spin) and extrinsic ($\phi_c$ and $t_c$)
parameters that describe the source binary.
%
For a NSBH population uniformly distributed in spatial volume, the detection
rate is $\propto\Olap(h^{\tr},h_b)^3$. To maximize the detection rate, it is
therefore crucial for the model waveform manifold, containing $h_b$, to 
faithfully reproduce the manifold of true waveforms that contains $h^\tr$.

In this paper, we take $S_n(|f|)$ to be the \textit{zero-detuning high power} 
noise curve for aLIGO~\cite{aLIGONoiseCurve,aLIGOsensitivity}. 
The peak GW frequency for the lowest binary masses
that we consider, i.e. for $m_1+m_2\simeq 8.4M_\odot$, is $\sim 3$~kHz during
ringdown. We sample the waveforms at $8192$~Hz, preserving the information 
content up to the Nyquist frequency $f_\mathrm{Ny}=4096$~Hz.

Our numerical waveforms begin at relatively low frequencies 
(between about $60\mbox{ Hz}$ and $80\mbox{ Hz}$), but nevertheless they do 
not completely span the detector's sensitive frequency band.
The discontinuity at the start of the NR 
waveform, because of Gibbs phenomena~\cite{WilbrahamGibbsJunk1848}, corrupts 
the Fourier transform. We therefore taper the start of the waveforms 
using a cosine tapering 
window, whose width is chosen to control the corruption of the resulting 
waveform in a way that the mismatches because of tapering stay below $0.2\%$. 
Additionally, to minimize residual spectral leakage, 
we apply an eighth order Butterworth high-pass filter with the 
cutoff 
frequency equal to the frequency of the waveform at the end of the tapering 
window. 
In Sec.~\ref{s2:nrerrorimpact}, we measure the effect of waveform 
conditioning on the NR waveforms by
comparing identically conditioned waveforms against
unconditioned-but-significantly-longer SEOBNRv2 waveforms.
\section{Comparison of waveform models}\label{s1:modelcomparison}
%
%
\begin{figure}
\centering
\includegraphics[width=\columnwidth]{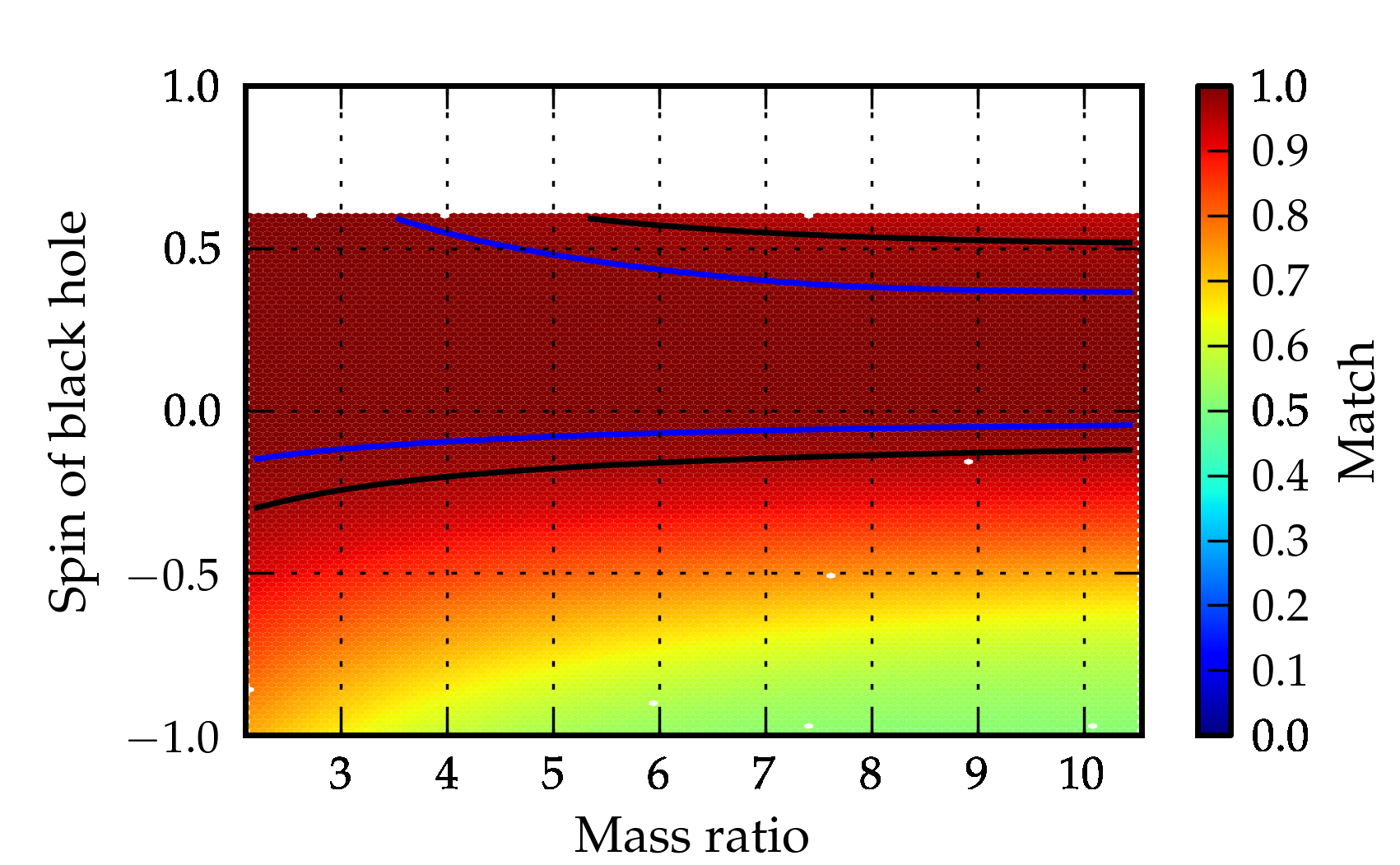}
\caption{In this figure, we show the match between the two SEOBNR models
we consider in this paper, as a function of the BH spin and binary
mass ratio. The mass of the NS is fixed at $1.4M_\odot$, and its spin is set to
$0$, consistent with the observed astrophysical population of 
NSs~\cite{2013ApJ...778...66K}. The blue (black) 
solid lines are the contours of $99\%$ ($97\%$) match. We find that
the matches between
the models drop sharply for $\chi_\mathrm{BH}\leq -0.15$.
}
\label{fig:seobv1v2_faith}
\end{figure}
%
%
\begin{figure*}
\centering

\includegraphics[width=\columnwidth]{%
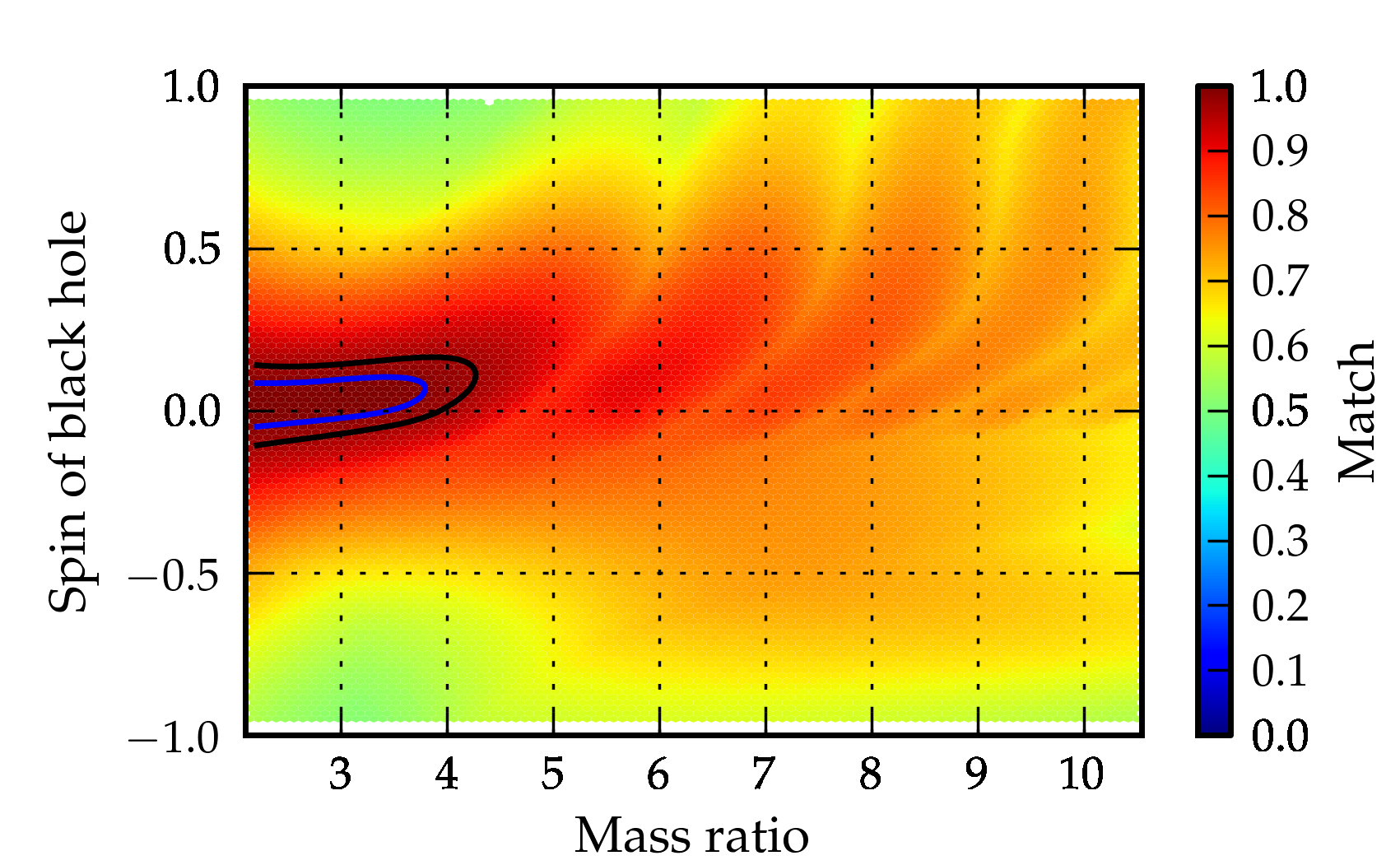}
\includegraphics[width=1.0\columnwidth]{%
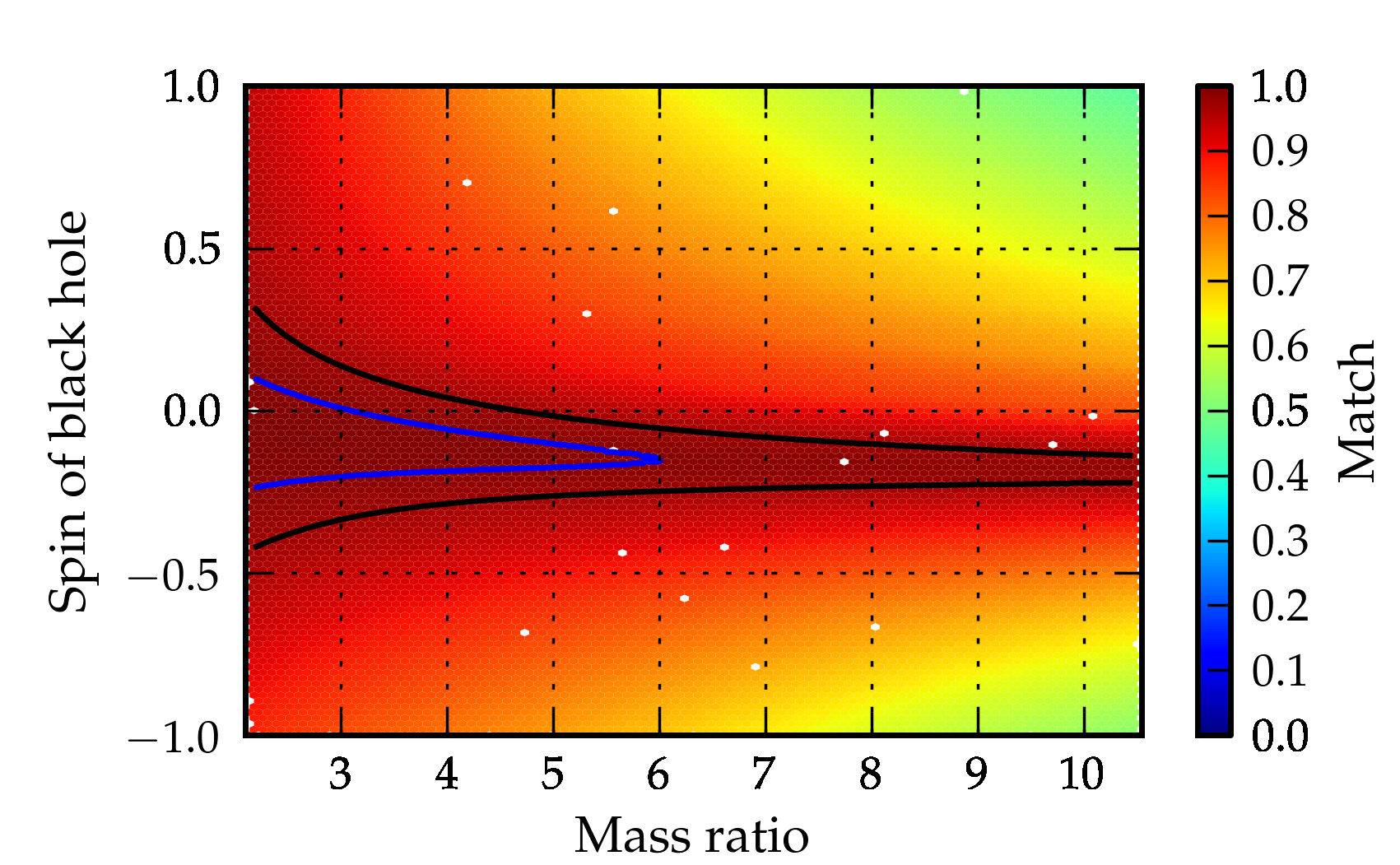}
\caption{
These plots are similar to Fig.~\ref{fig:seobv1v2_faith}, except 
they compare SEOBNRv2 with 
PhenomC (left panel) and TaylorF2 (right panel).
From the left panel, we observe that PhenomC and SEOBNRv2
are faithful to each other 
for very small BH spins and $q=m_1/m_2\lesssim 4$. 
Their matches drop sharply for moderately spinning binaries and also above 
moderate mass-ratios. Also, the fall in matches of PhenomC with increasing 
mass-ratio is not monotonic.
In the right panel, we find good agreement between TaylorF2 and SEOBNRv2
for binaries with very small BH spin. The best agreement, however, is for  
small anti-aligned spins, with the two models having $\lesssim 95\%$
matches for non-spinning binaries. Note that PhenomC uses TaylorF2 phasing
prescription the early inspiral, including, however, only the leading and 
next-to-leading order spin-dependent terms.
}
\label{fig:seobv2phenomcf2_faith}
\end{figure*}
\begin{figure}
\centering

\includegraphics[width=\columnwidth]{%
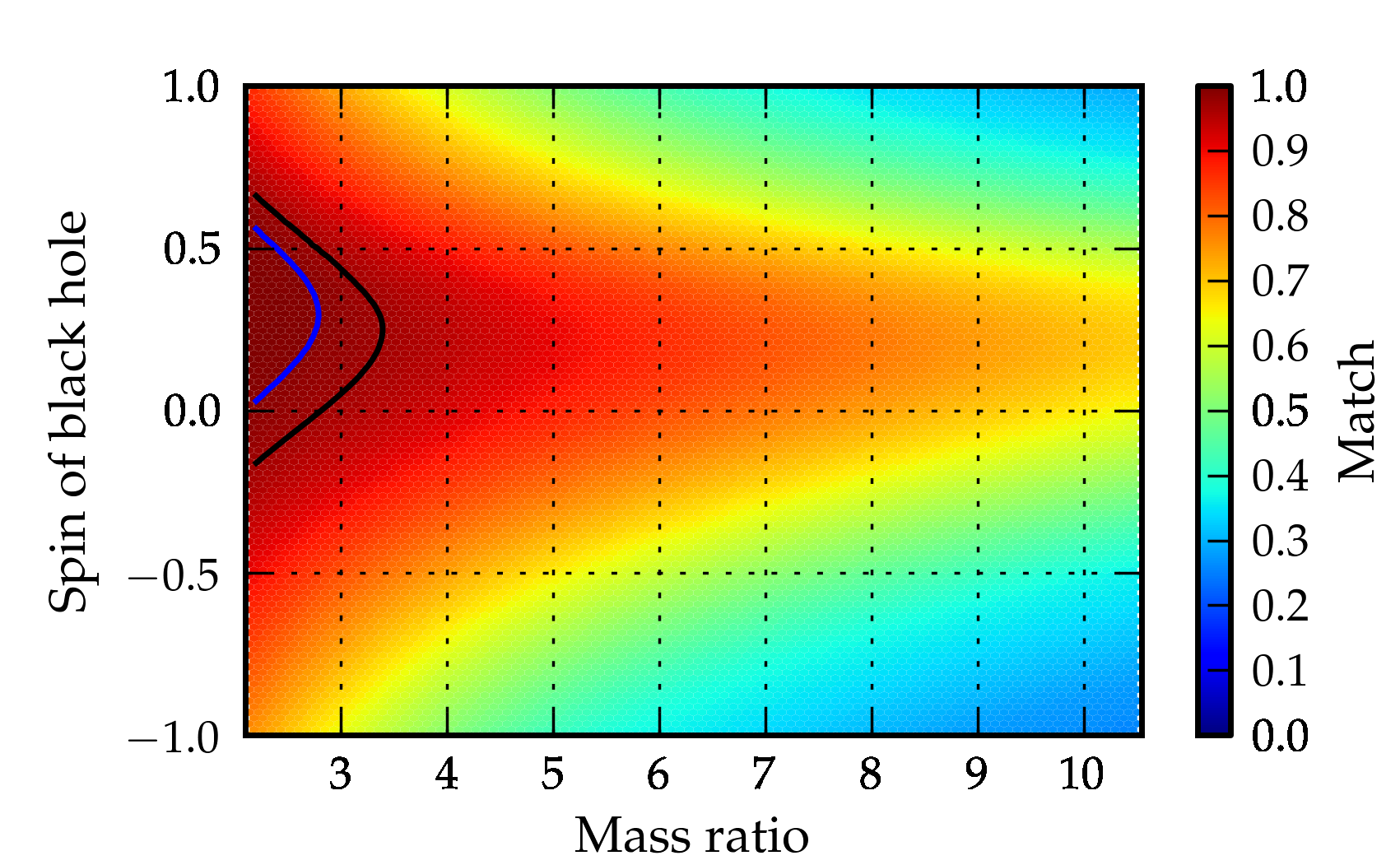}
\caption{
This figure is similar to Fig.~\ref{fig:seobv1v2_faith}, except it compares  
TaylorT4 with SEOBNRv2. We find that for mass ratios $q\geq 3.5$ or BH spin
$\chi_\mathrm{BH}\notin[-0.1,0.6]$, the two models disagree significantly, with
matches falling below $0.9$, down to $0.4$. 
}
\label{fig:seobv2t4_faith}
\end{figure}
\begin{figure*}
\centering
    \includegraphics[width=.7\columnwidth]{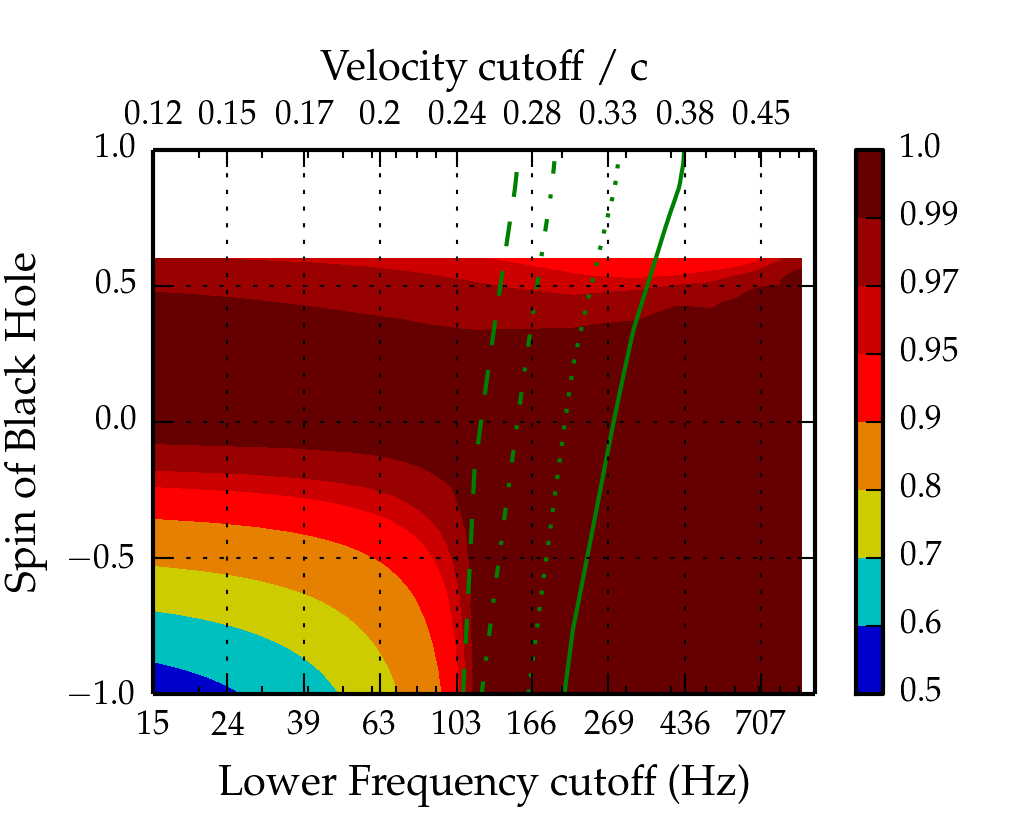}
    \includegraphics[width=.7\columnwidth]{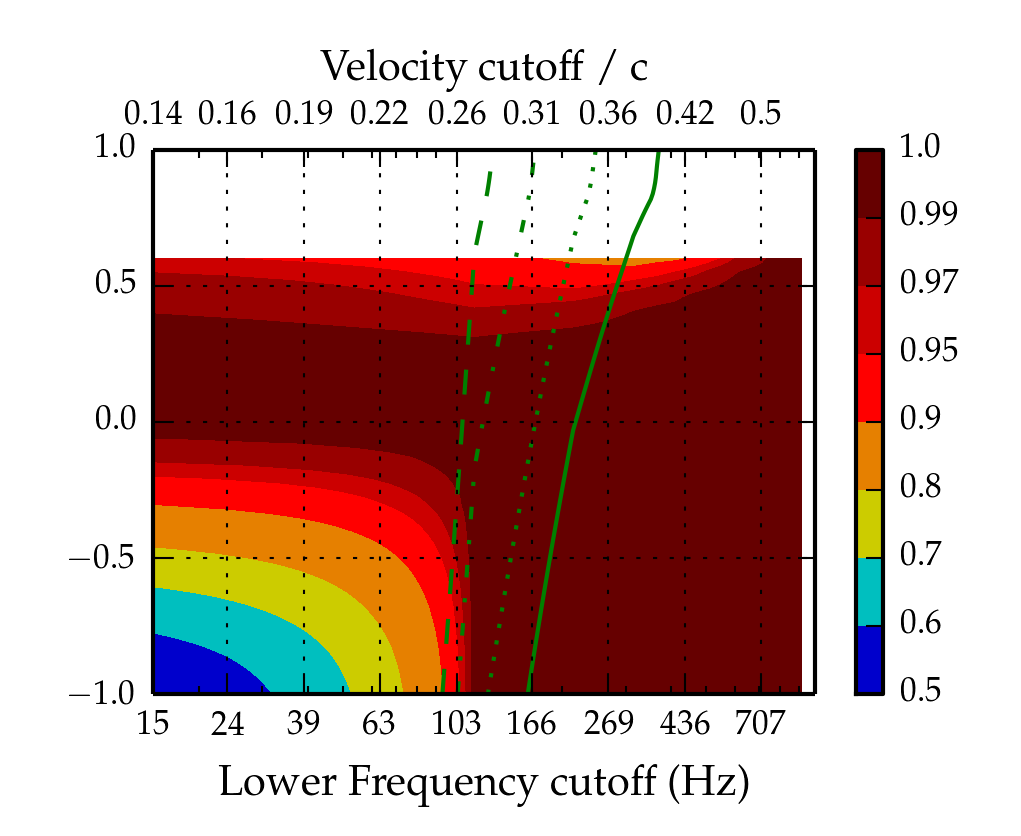}
    \includegraphics[width=.7\columnwidth]{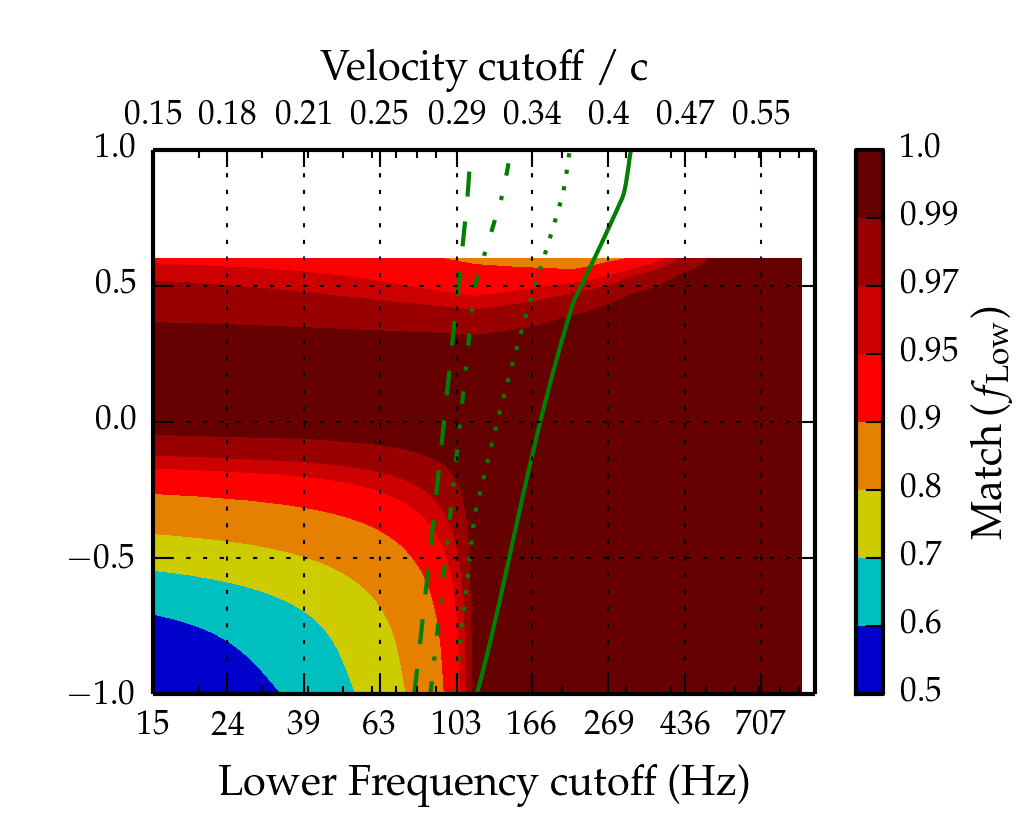}
\caption{
In this figure, we show the accumulation of mismatch between the two 
SEOBNR models that we consider in this paper by plotting the match 
(shown by colors) as a function of the low-frequency cutoff in the match
calculation (horizontal axes) and black-hole spins (vertical axes). 
  We choose $3$ sets of NSBH systems
  corresponding to $q=\{5,7,10\}$ and $m_\mathrm{BH}=\{7M_\odot,
  9.8M_\odot, 14M_\odot\}$ (from left to right panels,
  respectively). We vary the black-hole spin (vertical axes) over the
  validity range for SEOBNRv1~\cite{Taracchini:2012ig},
  $\chi_\mathrm{BH}\in [-1,0.6]$.  We fix the mass of the neutron star
  to $1.4M_\odot$ and the spin of the neutron star to zero. 
   The solid, dotted, dash-dotted and dashed lines
  are contours of the frequencies starting from which the binary
  merges after $5,10,20,30$ orbits, respectively. For
  anti-aligned BH spins, the matches are high when we integrate over
  the last few tens of orbits, but they fall
  \textit{significantly} as we include more orbits (lower frequencies)
  in the computation.  For moderate aligned BH spins, we find
  dephasing in the late-inspiral which is compensated for by lower
  frequency orbits.
  For comparison, the
  longest Numerical Relativity simulation to which either of the two
  models considered here have been calibrated to spans about $33$
  orbits~\cite{Taracchini:2013rva,sxscatalog}.  }
\label{fig:seobv1v2_freq}
\end{figure*}
\begin{figure*}
\centering
    \includegraphics[width=0.7\columnwidth]{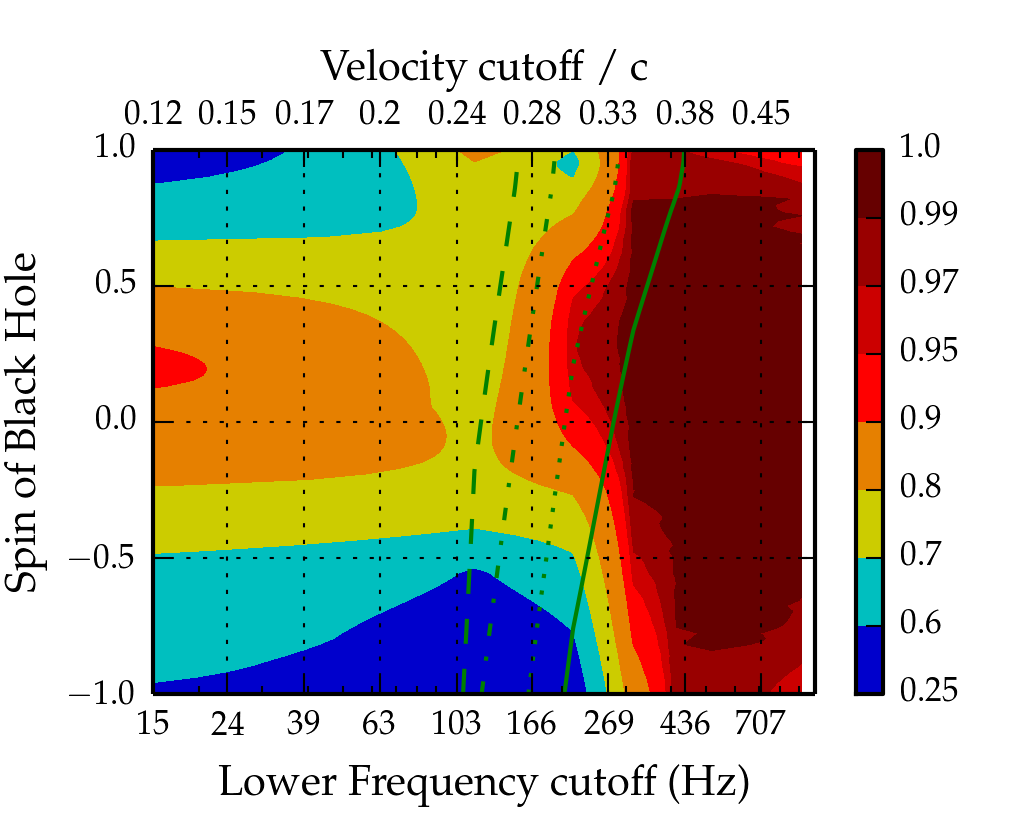}
    \includegraphics[width=0.7\columnwidth]{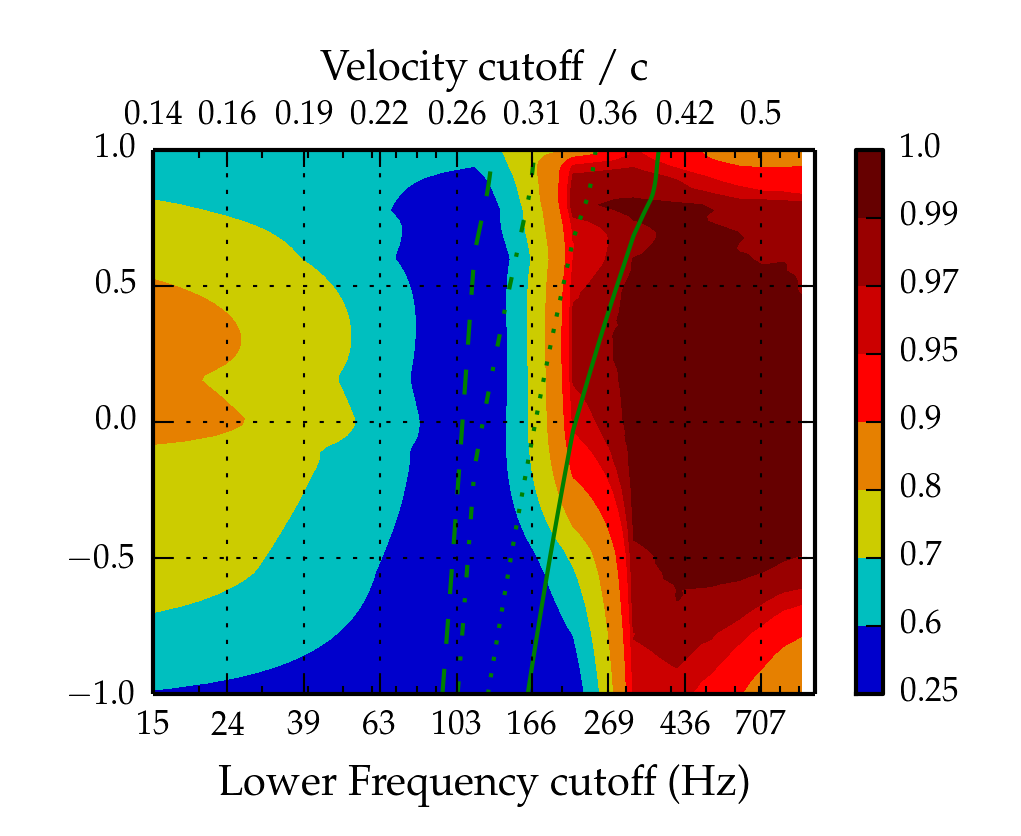}
    \includegraphics[width=0.7\columnwidth]{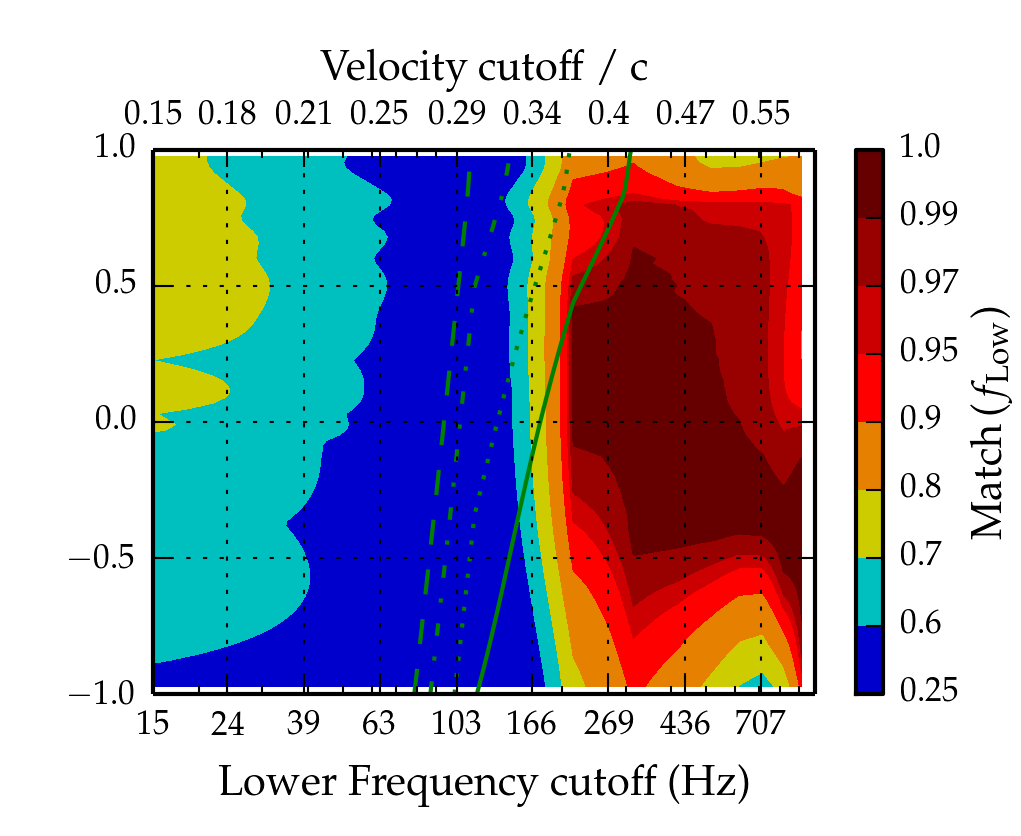}
\caption{These figures are similar to Fig.~\ref{fig:seobv1v2_freq}, 
with the only difference
that we compare here the PhenomC and SEOBNRv2 models. 
We observe that the two models agree over the last few ($< 10$) orbits, but
their matches drop sharply over earlier late-inspiral orbits. The inclusion
of very low-frequency orbits in match calculations leads to an increase in 
matches for $|\chi_\mathrm{BH}|\leq 0.2$. 
The pattern shown in these figures suggests that dephasing is accrued
rapidly close to the point where the model switches from TaylorF2 to its 
pre-merger phasing prescription.
}
\label{fig:seobv2phenomC_freq}
\end{figure*}
%
\begin{figure*}
\centering
    \includegraphics[width=0.7\columnwidth]{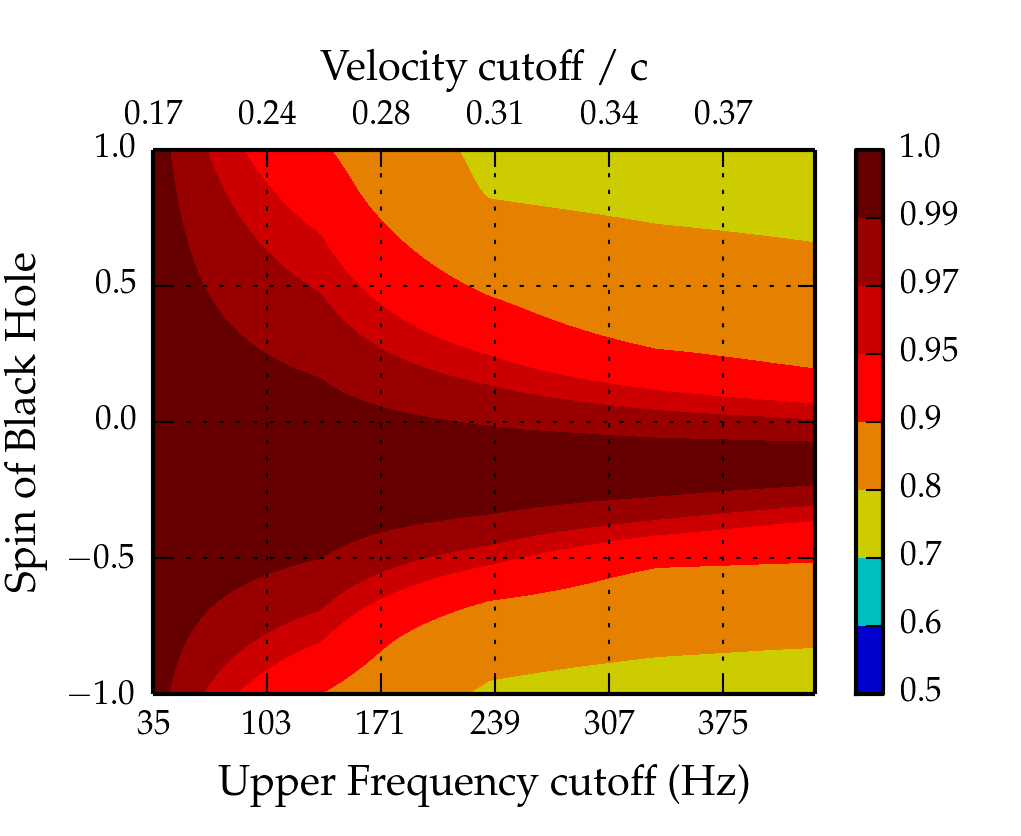}
    \includegraphics[width=0.7\columnwidth]{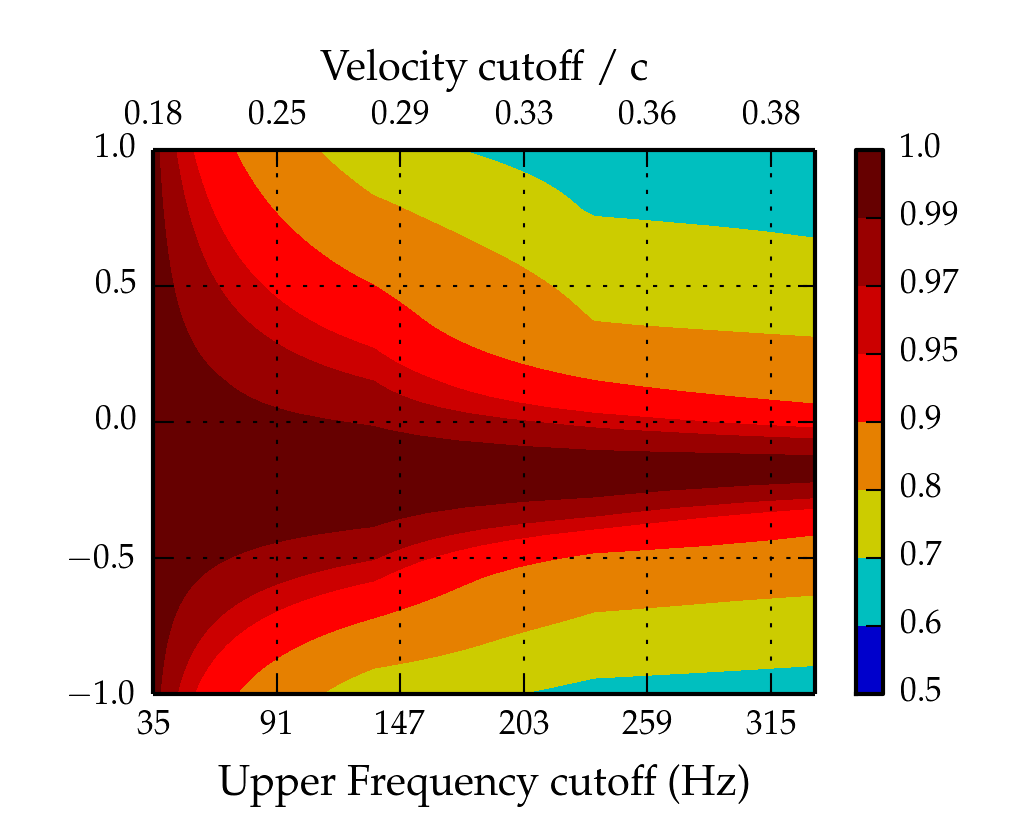}
    \includegraphics[width=0.7\columnwidth]{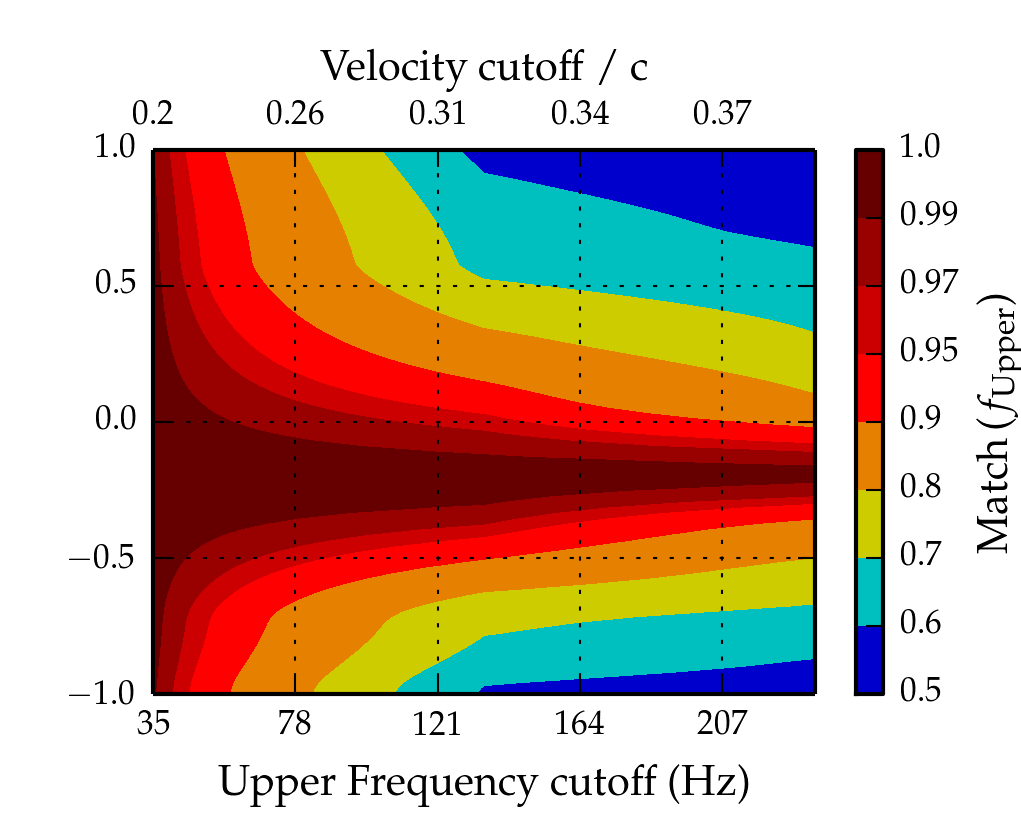}
\caption{These figures are similar to Fig.~\ref{fig:seobv1v2_freq} with two 
differences: (a) here we compare the TaylorF2 and SEOBNRv2 models, and (b)
the quantity shown is the match between the models as a function of the 
upper-frequency cutoff, with the low-frequency cutoff fixed at $15$~Hz.
Similar to the trend observed in Ref.~\cite{Nitz:2013mxa} for SEOBNRv1, 
TaylorF2 agrees with the SEOBNRv2 model at low frequencies, but their 
agreement drops \textit{significantly} during and after late-inspiral. 
The matches drop starting at lower frequencies with increasing BH spin 
magnitudes. These patterns are consistent with the fact that PN results 
decrease in accuracy with increasing orbital frequencies, especially as 
component spins becomes large.
}
\label{fig:seobv2f2_freq}
\end{figure*}
\begin{figure*}
\centering
    \includegraphics[width=0.7\columnwidth]{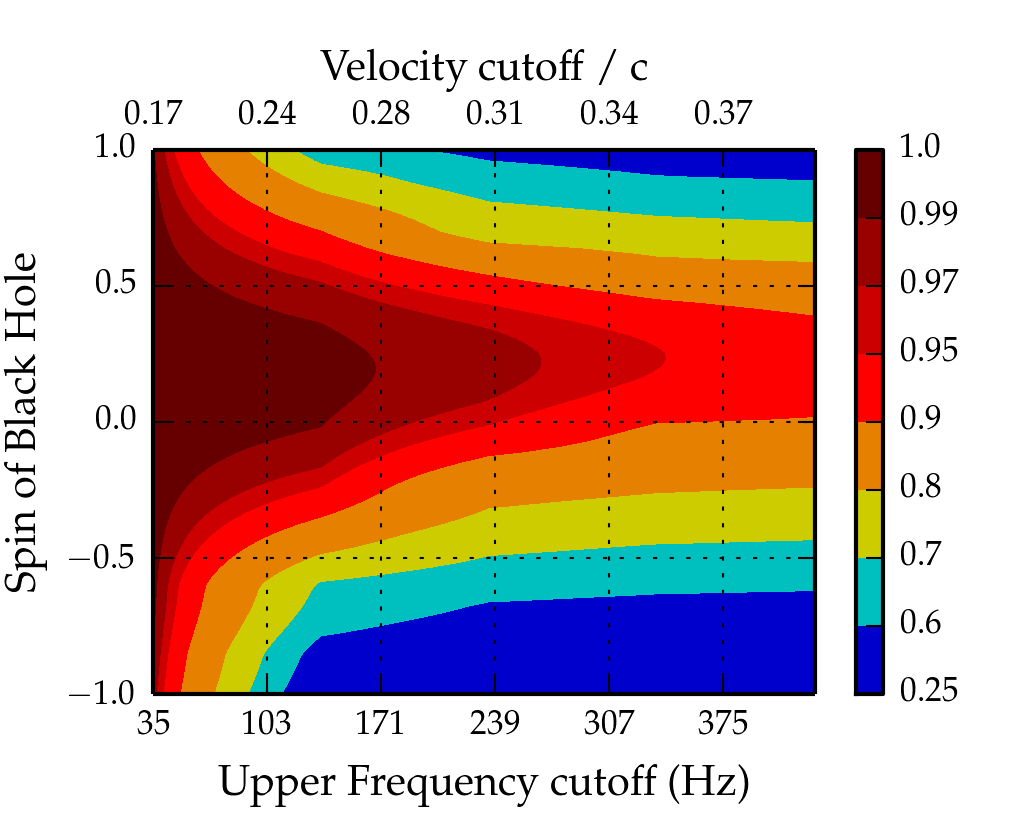}
    \includegraphics[width=0.7\columnwidth]{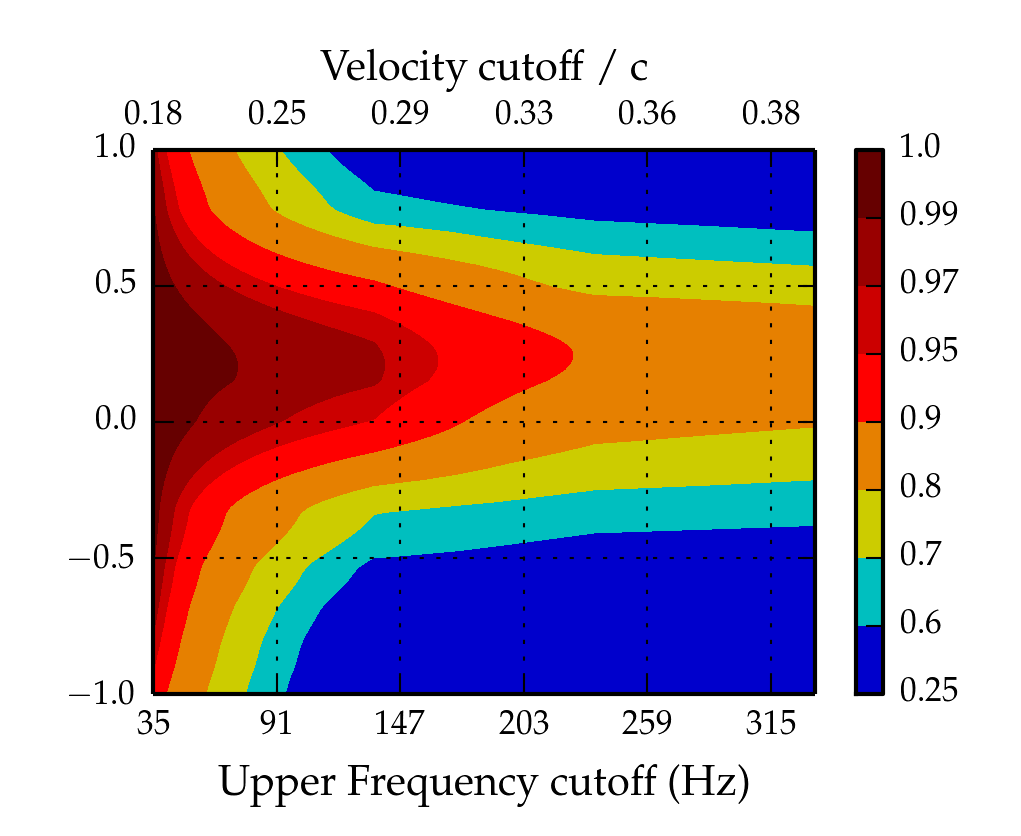}
    \includegraphics[width=0.7\columnwidth]{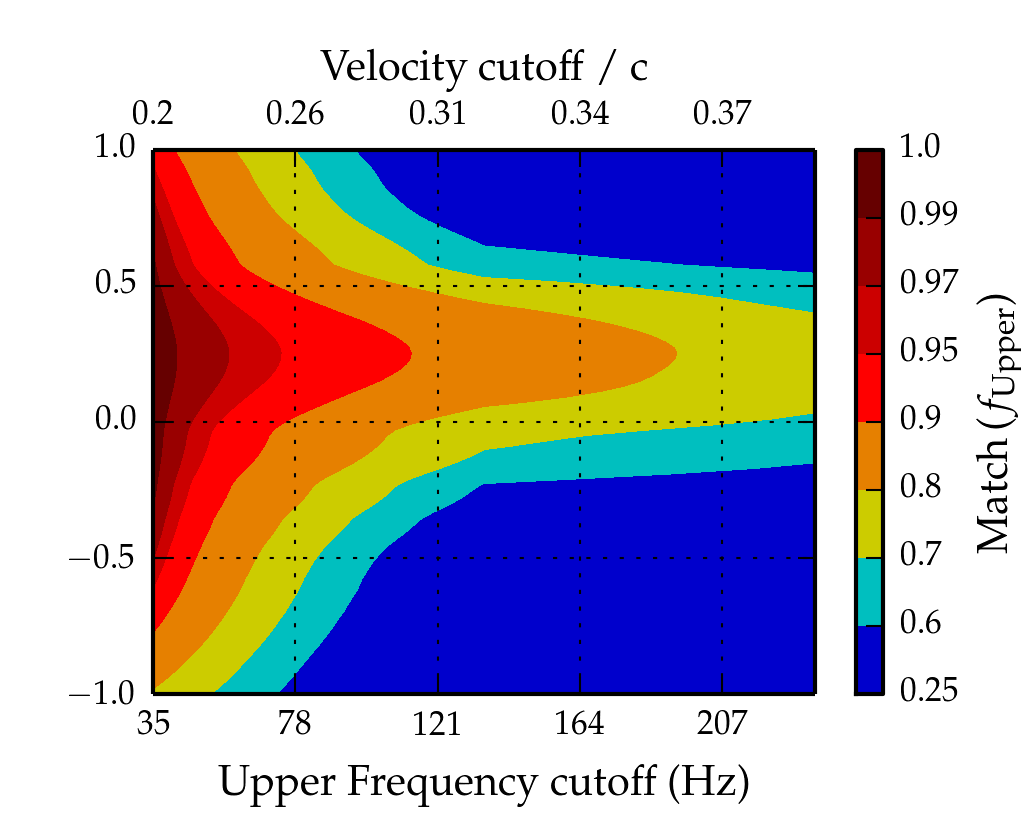}
\caption{These figures are similar to Fig.~\ref{fig:seobv2f2_freq} with the only 
difference that here we compare the TaylorT4 and SEOBNRv2 model.
Comparing to Fig.~\ref{fig:seobv2f2_freq}, we find that (a) the patterns of 
mismatch accumulation in these figures are qualitatively similar to TaylorF2, 
and 
(b) the disagreement between TaylorT4 and SEOBNRv2 accumulates starting at
\textit{lower} frequencies and \textit{more drastically} compared to TaylorF2.
These results suggest that higher order PN terms in orbital phasing, 
especially spin dependent terms, are required to obtain a more accurate PN 
description of the late-inspiral phase.
}
\label{fig:seobv2t4_freq}
\end{figure*}
%

In this section, we show the faithfulness between waveforms from different 
approximants, where we choose the physical parameters to be consistent with 
NSBH sources. We compare the inspiral-merger-ringdown (IMR) models SEOBNRv1
and PhenomC, and the PN models TaylorT4 and TaylorF2, with the SEOBNRv2
model.
We also show how the disagreement between approximants builds up over the 
coarse of a binary's inspiral, by computing their faithfulness over different
GW frequency intervals.
For both, we take SEOBNRv2 as the fiducial model because it has been calibrated
against the highest number of high-accuracy NR simulations of aligned-spin 
binaries ($38$ in total, with mass ratio up to $\sim 8$), and is therefore
likely to be the most accurate representation of true waveforms available at
present).
%
%
Overall, we find that (i) the two SEOBNR models (SEOBNRv1,2) disagree 
significantly for anti-aligned spinning binaries (matches below $80\%$), with
their mismatches accumulating over lower frequency inspiral orbits;
(ii) PhenomC and SEOBNRv2 produce drastically different waveforms over most
of the NSBH parameter space, except for the small mass-ratio $+$ small spin corner;
and (iii) both PN models show slightly better agreement with SEOBNRv2 than 
PhenomC, but still restricted to small mass-ratios and small component spins,
which is consistent with~\cite{Nitz:2013mxa}.
%

\subsection{Faithfulness of models}\label{s2:faithfulness}

In Fig.~\ref{fig:seobv1v2_faith}, we examine the faithfulness of the
two SEOBNR models. Neither of these models were used as templates in
past LIGO searches, because they were published after initial LIGO 
searches were completed, but both are promising
candidate models for aLIGO. Focusing on stellar-mass NSBH binaries, we
fix the NS mass to $1.4M_\odot$, the NS spin to $0$, and allow the BH
mass to vary over $[3,15]M_\odot$ and the BH spin to vary over the allowed 
range of SEOBNRv1 $[-1,0.6]$~\cite{Taracchini:2012ig}. We see that the
agreement between the models is primarily influenced by the BH spin
and secondarily by the mass ratio. As expected, both agree in the
comparable mass and non-spinning limits, where both incorporate
information from NR simulations. We also find good agreement for
\textit{aligned} BH spins. However, when the BH spin is
\textit{anti-aligned} with the orbital angular momentum, SEOBNRv1 and
SEOBNRv2 produce \textit{significantly} different waveforms, with
matches dropping below $0.8$ for $\chi_\mathrm{BH}\leq -0.5$. This
demonstrates that the more recent SEOBNRv2 model incorporates
different spin-dependent phasing terms. However, to make statements
about the \textit{accuracies} of either, we must analyze both using
high-accuracy NR waveforms, a comparison we turn to in the next section.

In Fig.~\ref{fig:seobv2phenomcf2_faith}, we show the faithfulness of
the PhenomC (left panel) and TaylorF2 (right panel) models against
SEOBNRv2. We notice that PhenomC agrees with SEOBNRv2 \textit{only}
for very mildly spinning binaries with $|\chi_\mathrm{BH}|\lesssim
0.1$ \textit{and} comparable mass ratios $q\lesssim 4$. Over the
remainder of the parameter space, the matches between PhenomC and
SEOBNRv2 were found to be low.  This is somewhat surprising, since 
PhenomC has been calibrated against spinning NR simulations with
$q$ up to $4$ as well, albeit spanning fewer orbits and produced using
a different NR code~\cite{Bruegmann:2006at,Husa:2007rh}.
Comparing with the right panel of
Fig.~\ref{fig:seobv2phenomcf2_faith}, we find that TaylorF2 agrees
with SEOBNRv2 more closely for rapidly spinning NSBH systems with
$q\lesssim 3$, as well as for binaries with small BH spin magnitude at
all mass ratios. Since the inspiral portion of PhenomC phasing is the
same as TaylorF2 (with the caveat that the former does not include the
recently published spin-dependent $3$~PN and $3.5$~PN
contributions~\cite{Blanchet:2012sm, Bohe:2012mr}), we conclude that
their differences arise from the post-merger phasing prescription of
PhenomC.  We study this further later in this section, where we 
highlight the phase of binary coalescence where different approximants
disagree.

We further show the faithfulness between TaylorT4 and SEOBNRv2 models
in Fig.~\ref{fig:seobv2t4_faith}. We find that their faithfulness
drops sharply with increasing mass ratio, falling below $0.9$ for mass-ratios 
$q\gtrsim 4$ for all values of BH spin. \textit{Only} for near-equal mass 
low-spin binaries does TaylorT4 agree well with SEOBNRv2, which is consistent 
with past comparisons with NR simulations~(e.g., Fig.~7 
of~\cite{MacDonald:2012mp}). Comparing with the
right panel of Fig.~\ref{fig:seobv2phenomcf2_faith}, we see that (a) TaylorF2
has better overall agreement with SEOBNRv2, and (b) TaylorF2 agrees better for
small anti-aligned spins and TaylorT4 for small aligned spins. These differences
are expected to decrease in future PN approximants, as higher order PN terms
become available.
\subsection{Accumulation of mismatches with frequency}\label{s2:mismatchaccumulation}

For the adiabatic early-inspiral phase where the binary is well separated and 
inspirals relatively slowly, all GW models considered here are based on PN
results. 
As the binary tightens, the PN approximation becomes less
accurate. In order to capture the late-inspiral/plunge and merger phases,
the IMR models either use purely phenomenological prescriptions or re-sum 
truncated PN results to add terms at all unknown higher orders in a controlled 
way. In both approaches, the model is calibrated against NR 
simulations of binary mergers, and models which are more 
extensively calibrated tend to be more robust. 
But do the different models agree in the late-inspiral 
regime, where the PN approximation is not valid and where we have no NR 
simulations available? To answer this question, we study here the 
mutual disagreement between models over different phases of binary coalescence.


First, we examine the mismatch accumulation between the two EOB models 
for three representative sets of NSBH masses. In
Fig.~\ref{fig:seobv1v2_freq}, the three panels correspond to mass-ratios 
$q=m_1/m_2=\{5,7,10\}$ (left to right), with the NS mass fixed at $1.4M_\odot$.
The color shows the match between SEOBNRv1 and SEOBNRv2 as a 
function of the lower frequency cutoff on the match integral 
(shown on the x axes), for different values of BH spin (shown on the y axes).
The four green curves on each panel are contours of the frequencies that mark 
``N orbits to merger'', with $N=30,20,10,5,$ respectively from left to right.
We first note that the two models agree well for non-spinning
binaries. When the BH spin is aligned to the orbital momentum, we
observe some dephasing for binaries with $\chi_\mathrm{BH}\gtrsim 0.3$,
that accumulates over the last $30$ or so orbits. When integrated over the 
entire waveform, this dephasing gets compensated for by the lower freqeuency
orbits.
For anti-aligned BH spins, however, the agreement between the two
models is \textit{significantly} low during the early inspiral. Over
the last $15-20$ orbits the two models have matches $>0.99$, but they
drop below $0.95$ around the $20-30$ orbit mark, and further decrease
\textit{monotonically} as lower frequencies are included in the match 
calculation. Therefore, for anti-aligned spins, we would expect that
NR simulations with lengths $\lesssim 30\,\mbox{orbits}$, as used 
by~Ref.~\cite{Taracchini:2013rva} for SEOBNRv2, would match well 
with both the SEOBNRv1 and SEOBNRv2 models, but that they would 
drastically disagree with at least one of the models early on.
%
%
Given the lack of clear convergence 
of the SEOBNR models, we will investigate their \textit{accuracy} in the 
following section by comparing them to long NR waveforms that probe the 
low-frequency regime where the models disagree.

Next, we compare the PhenomC and SEOBNRv2 models by computing their matches 
with varying lower frequency cutoffs. The results are shown in 
Fig.~\ref{fig:seobv2phenomC_freq}, where all three panels are similar to 
Fig.~\ref{fig:seobv1v2_freq} with the only difference that SEOBNRv1 is replaced 
by PhenomC. 
We first note that the two models agree during the very early inspiral where 
PhenomC reduces to TaylorF2, as well as over the last few pre-merger orbits.
Most of their dephasing accumulates in a relatively narrow frequency range 
centered at $\sim 100$~Hz, which is where PhenomC switches to its 
phenomenological phasing prescription.
We also find that the spin on the BH affects the model agreement in two ways:
(i) their dephasing increases with spin magnitude, and (ii) it also increases
as the BH spin gets increasingly anti-aligned with the orbital angular momentum.
%
%


Inspiral-only PN models have been shown to agree with SEOBNRv1 during
early inspiral and to monotonically diverge as the orbital frequency
rises~\cite{Nitz:2013mxa}. To quantify their agreement with our
fiducial model, SEOBNRv2, we compute matches between PN and SEOBNRv2
as a function of the \textit{upper} frequency cutoff (with the lower
cutoff fixed at $15$~Hz).  In Fig.~\ref{fig:seobv2f2_freq} we show the
results for TaylorF2, with the upper frequency cutoff on the x-axes,
BH spin on the y-axes and colors showing matches. The three panels
correspond to the same representative NSBH systems as in
Fig.~\ref{fig:seobv1v2_freq}. We find that for mildly anti-aligned BH
spins with $\chi_\mathrm{BH}\gtrsim-0.2$, TaylorF2 agrees with SEOBNRv2
through most of late-inspiral. For other BH spin values, the two
models start disagreeing at relatively low frequencies, e.g., for a
$q=5$ binary with $\chi_\mathrm{BH}=-0.6$, the match drops to $0.9$
between $15$~Hz and $300$~Hz.
%
%
%
In Fig.~\ref{fig:seobv2t4_freq} we show the same results for TaylorT4. We find 
that for mild aligned BH spins with $\chi_\mathrm{BH}\lesssim +0.2$, TaylorT4 
agrees with SEOBNRv2 through a significant portion of the inspiral. For higher 
or anti-aligned BH spins, the matches fall sharply below $0.7$ as more of high 
frequency orbits are integrated over. The agreement gets restricted to even 
lower frequencies as the mass-ratio increases, for the entire range of BH 
spins. From this systematic frequency-dependent discrepancy between PN and 
SEOBNRv2, we expect that higher order spin-dependent PN corrections to orbital 
phasing are required for a better PN modeling of the late-inspiral waveform.

Finally, we note that some of the results presented in this section are
qualitatively similar to~\cite{Nitz:2013mxa}, with the differences that
(a) we additionally include the recently published spin-orbit tail 
(3PN)~\cite{Blanchet:2012sm} and the next-to-next-to-leading order spin-orbit
(3.5PN) contributions~\cite{Bohe:2012mr} in both of the PN models, and (b) we 
include the SEOBNRv2 and PhenomC models here, both of which are capable of
modeling binaries with very high black hole spins 
$\chi_\mathrm{BH}\simeq +1$.



\section{Comparison with Numerical Relativity 
simulations}\label{s1:numrelcomparison}
%
%
\begin{figure*}
\centering
\includegraphics[width=\columnwidth]{%
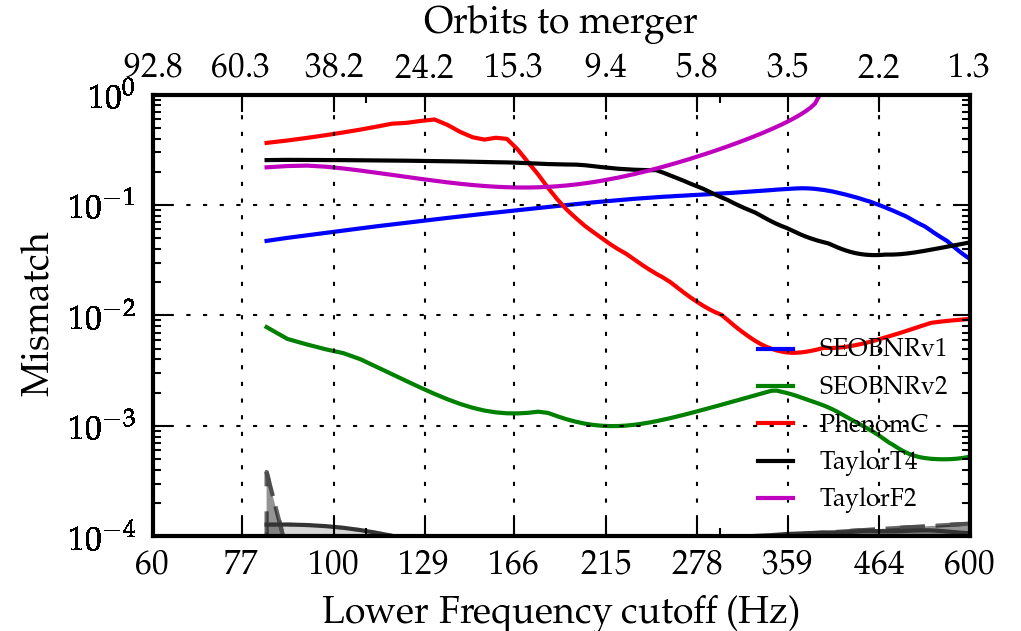}
\includegraphics[width=\columnwidth]{%
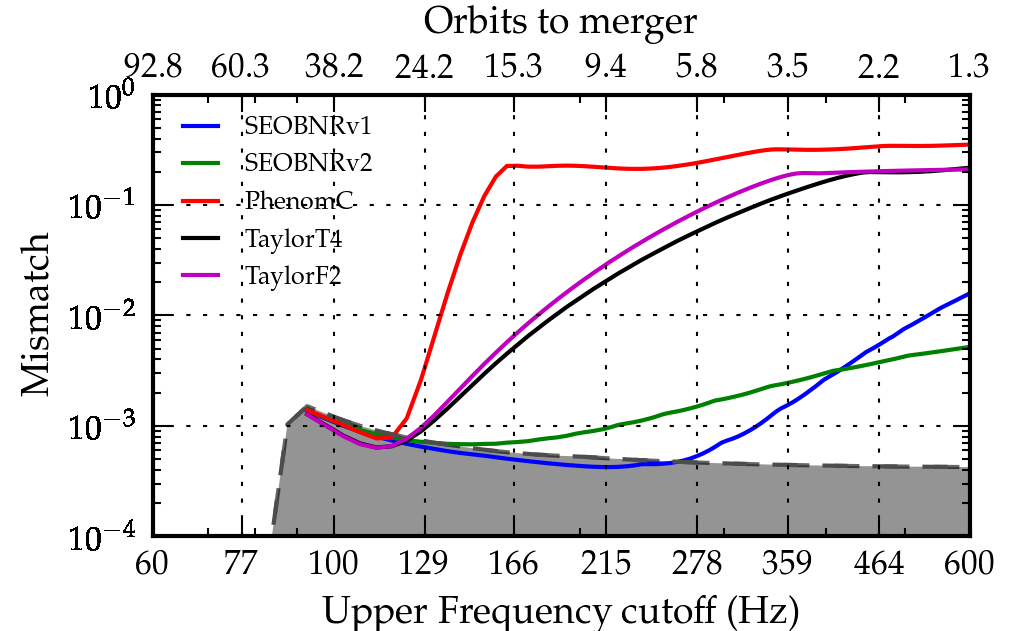}
\caption{For a NSBH binary with 
$q=m_\mathrm{BH}/m_\mathrm{NS}=9.8M_\odot/1.4M_\odot =7$ 
and $\chi_\mathrm{BH} = +0.6$, these figures show the mismatch of TaylorF2,
TaylorT4, SEOBNRv1, SEOBNRv2 and PhenomC 
waveforms against our simulation ID~SXS:BBH:202 (see Table~\ref{table:simlist}) as a 
function of the lower (left) and upper (right) frequency cutoff on the overlap 
integral. The NR waveform starts from GW
frequency $f_0 \simeq 80$~Hz.
The dashed curve with dark grey shading (in both panels) shows mismatches because of 
the tapering and high-pass filtering of NR waveforms, which we do to reduce 
Gibbs phenomena and spectral leakage upon Fourier transformation. Because of the 
width of the tapering window, we begin filtering at $82.5$~Hz.
The solid curve with light grey shading (only barely visible at the bottom 
of the left panel and below the range of mismatches shown in the  
right panel) shows 
the mismatches between NR waveforms at the highest and second-highest available 
numerical resolutions. 
}
\label{fig:model_nrmismatch_freq_low_s06}
\end{figure*}
\begin{figure*}
\centering    
\includegraphics[width=\columnwidth]{%
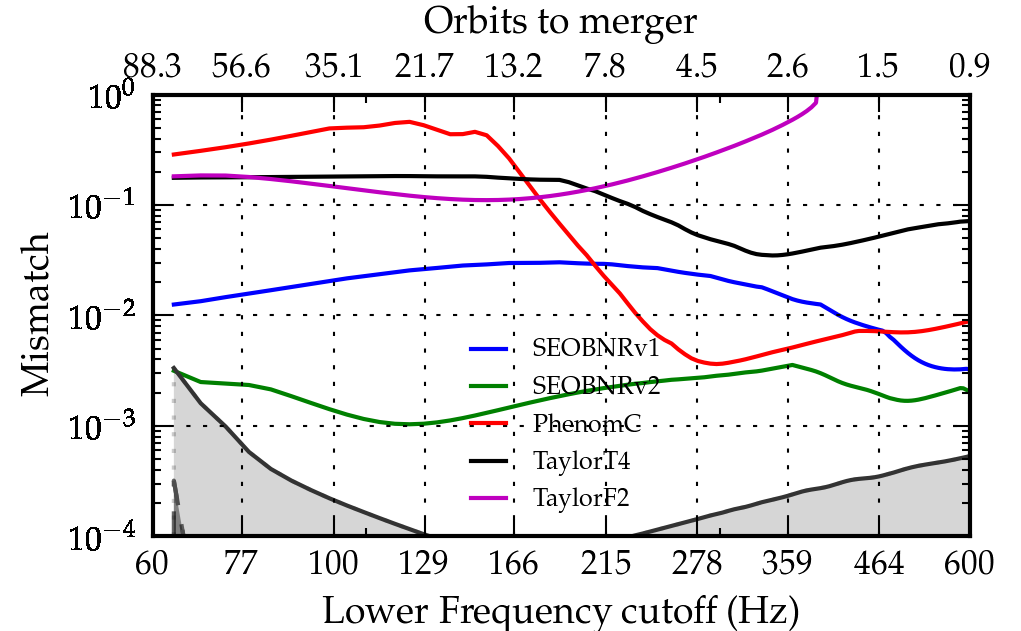}
\includegraphics[width=\columnwidth]{%
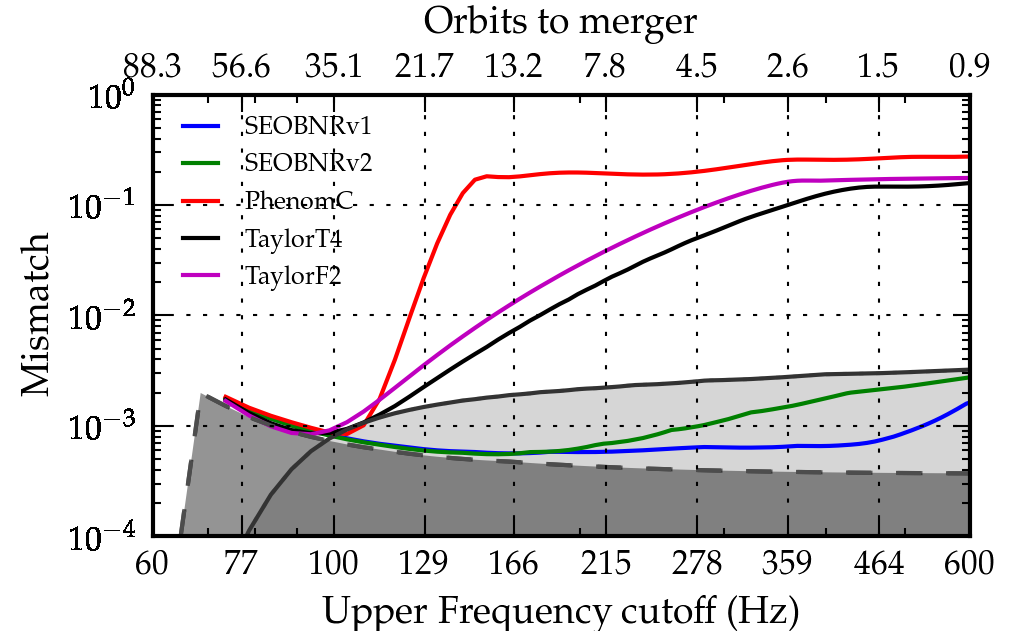}\\
\includegraphics[width=\columnwidth]{%
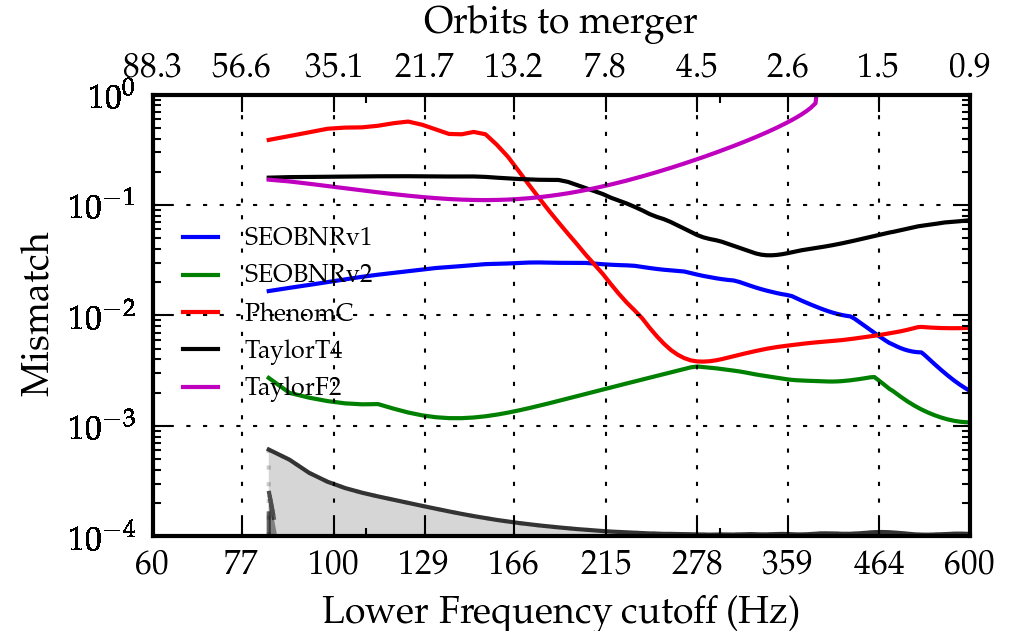}
\includegraphics[width=\columnwidth]{%
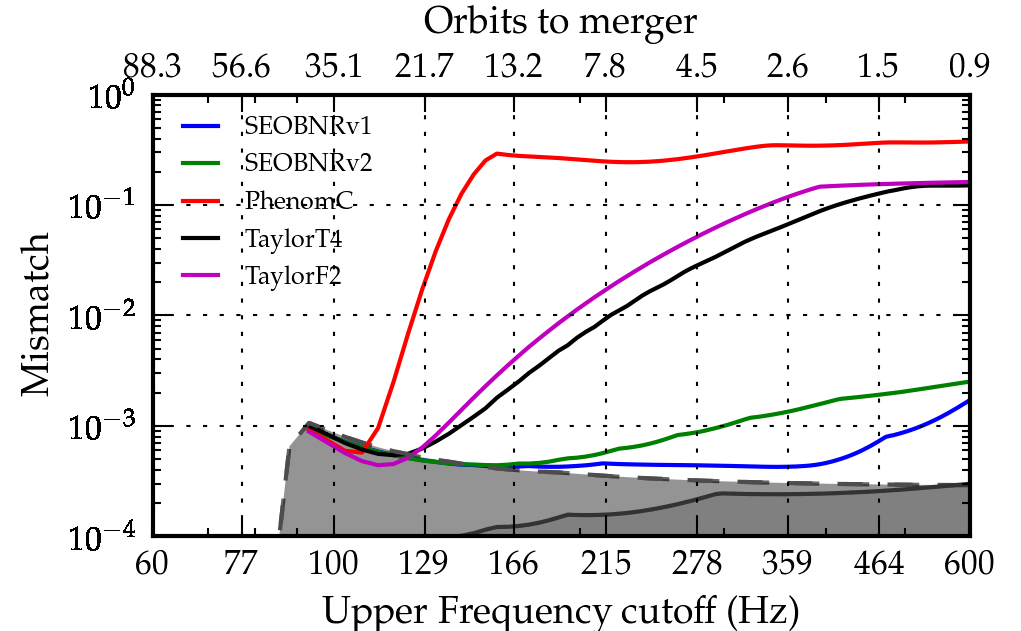}
\caption{The top two panels of this figure are similar to the two panels of 
Fig.~\ref{fig:model_nrmismatch_freq_low_s06}, with the difference that the 
system considered here has $q=m_\mathrm{BH}/m_\mathrm{NS}=9.8M_\odot/1.4M_\odot 
=7$, 
$\chi_\mathrm{BH} = +0.4$, and starts at GW frequency $\simeq 60$~Hz. This 
corresponds to simulation ID~SXS:BBH:204 (see Table~\ref{table:simlist}).
The bottom two panels are similar to the top two with the difference that these 
correspond to simulation ID~SXS:BBH:203, which has the same physical parameters
as ID~SXS:BBH:203 but a higher starting GW frequency, i.e. $\simeq 80$~Hz.
Because of the width of the tapering windows, we begin filtering at $63.5$~Hz and 
$83$~Hz, respectively.
}
\label{fig:model_nrmismatch_freq_low_s04}
\end{figure*}
\begin{figure*}
\centering
    
\includegraphics[width=\columnwidth]{%
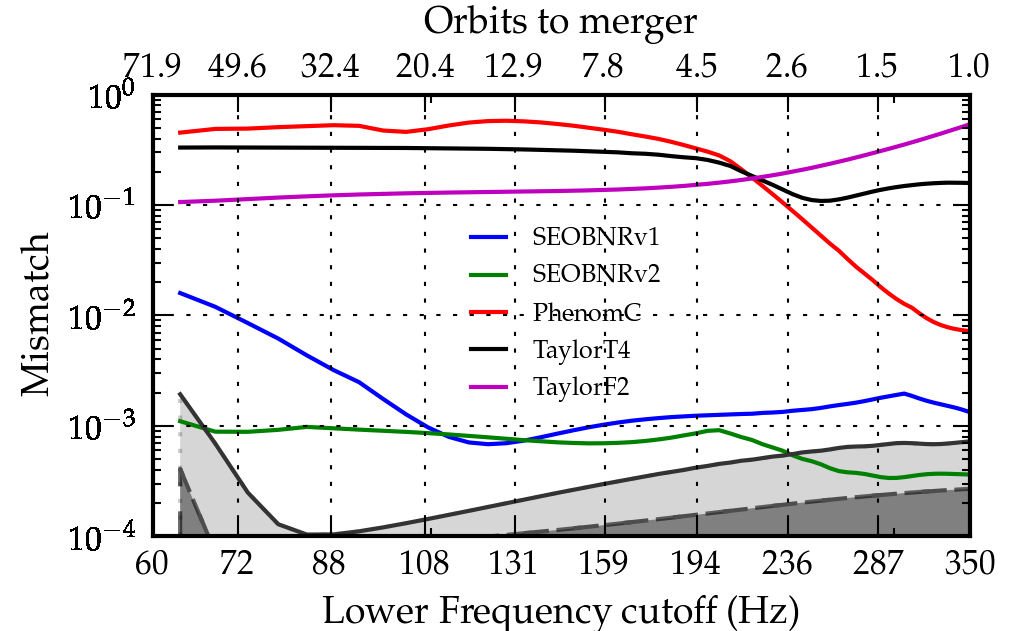}
\includegraphics[width=\columnwidth]{%
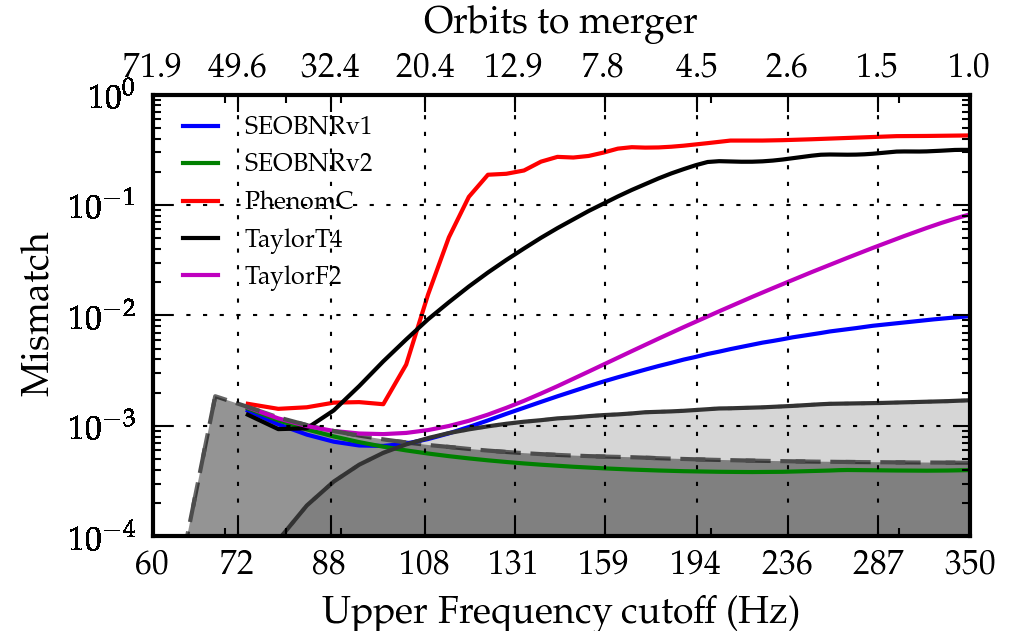}\\
\includegraphics[width=\columnwidth]{%
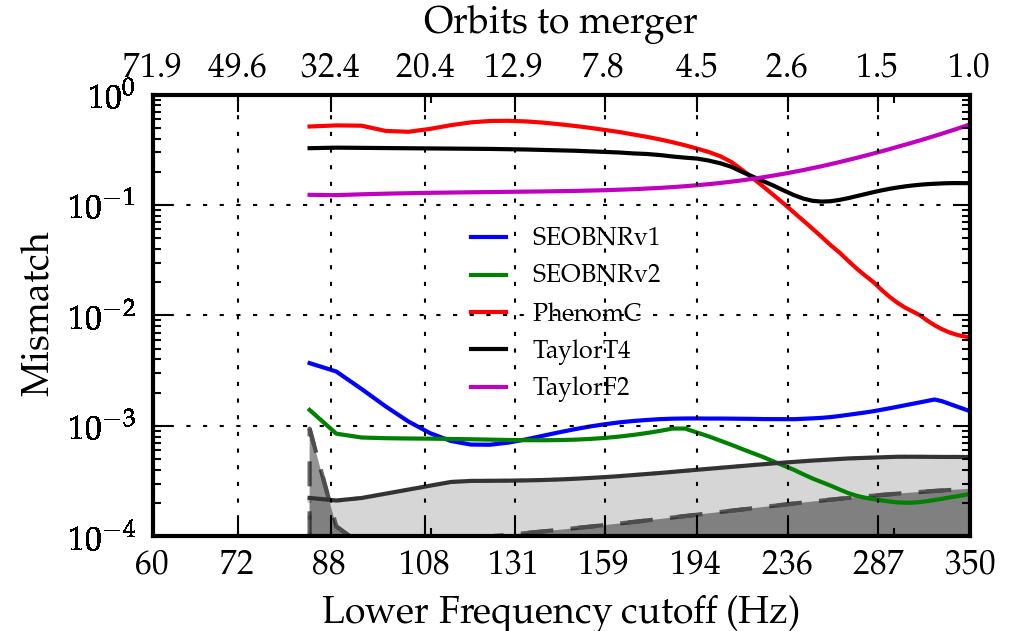}
\includegraphics[width=\columnwidth]{%
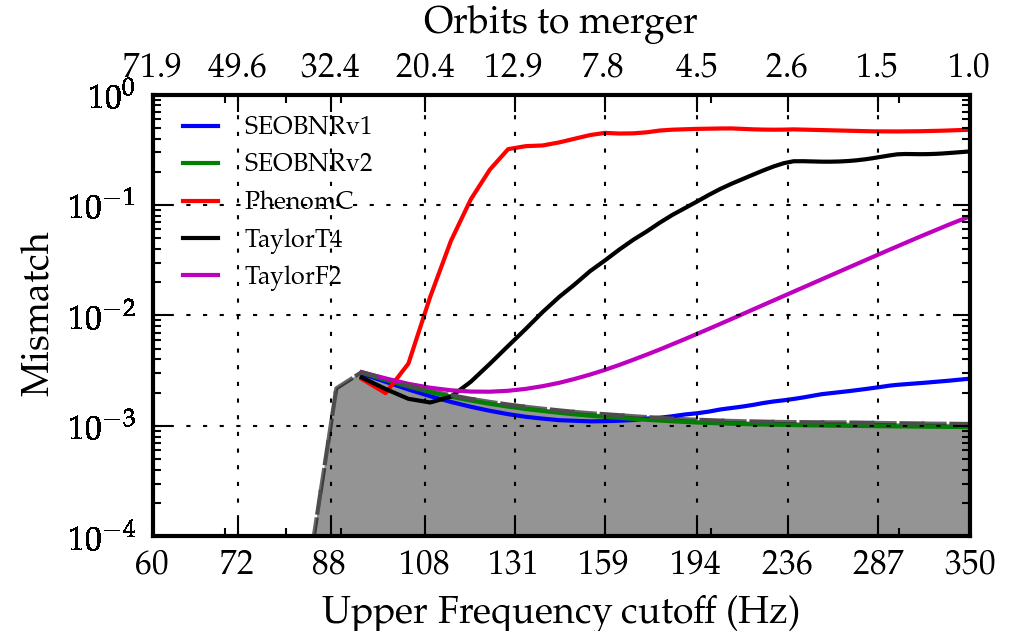}
\caption{The top two panels of this figure are similar to the two panels of 
Fig.~\ref{fig:model_nrmismatch_freq_low_s06}, with the difference that the 
system considered here has 
$q=m_\mathrm{BH}/m_\mathrm{NS}=9.8M_\odot/1.4M_\odot =7$, 
$\chi_\mathrm{BH} = -0.4$, and starts at GW frequency $\simeq 60$~Hz. This 
corresponds to simulation ID~SXS:BBH:206 (see Table~\ref{table:simlist}).
The bottom two panels are similar to the top two with the difference that these 
correspond to simulation ID~SXS:BBH:205, which has the same physical parameters as 
ID~SXS:BBH:206 but a higher starting GW frequency, i.e. $\simeq 80$~Hz.
Because of the width of the tapering windows, we begin filtering at $63.5$~Hz and 
$84$~Hz, respectively.
}
\label{fig:model_nrmismatch_freq_low_s-04}
\end{figure*}
\begin{figure*}
\centering
    
\includegraphics[width=\columnwidth]{%
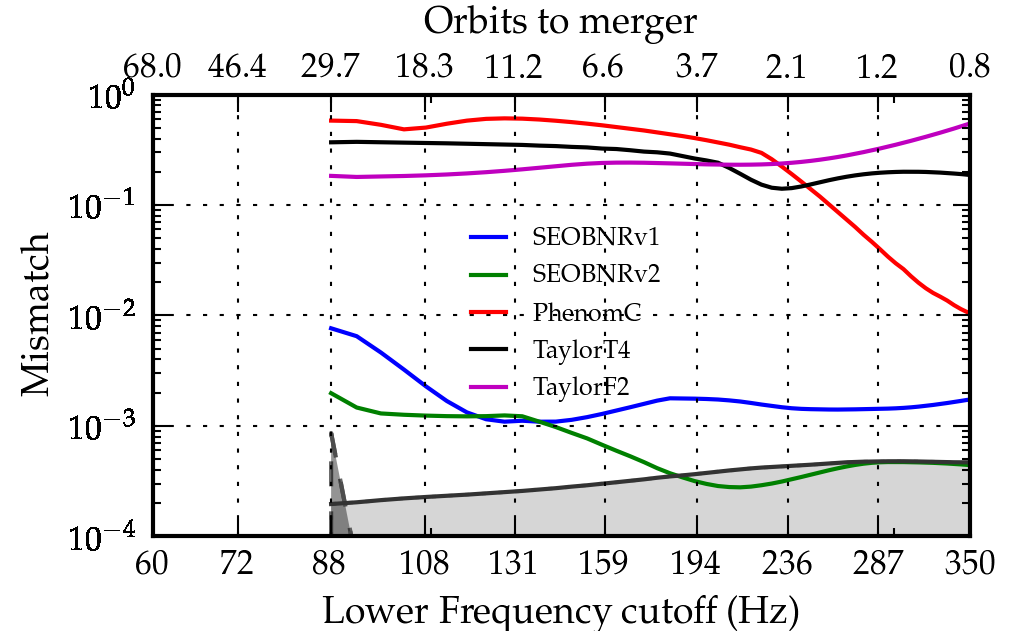}
\includegraphics[width=\columnwidth]{%
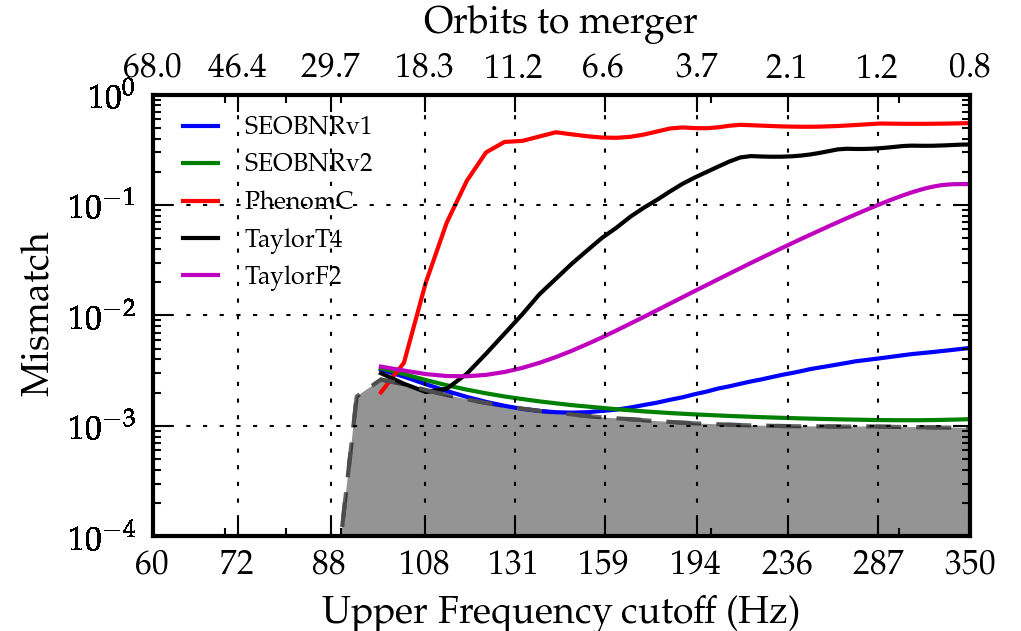}
\caption{These figures are similar to the two panels of 
Fig.~\ref{fig:model_nrmismatch_freq_low_s06}, with the difference that the 
system considered here has 
$q=m_\mathrm{BH}/m_\mathrm{NS}=9.8M_\odot/1.4M_\odot =7$, 
$\chi_\mathrm{BH} = -0.6$, and starts at GW frequency $\simeq 80$~Hz. This 
corresponds to simulation ID~SXS:BBH:207 (see Table~\ref{table:simlist}).
Because of the width of the tapering window, we begin filtering at $88$~Hz.
}
\label{fig:model_nrmismatch_freq_low_s-06}
\end{figure*}
\begin{figure*}
\centering

\includegraphics[width=\columnwidth]{%
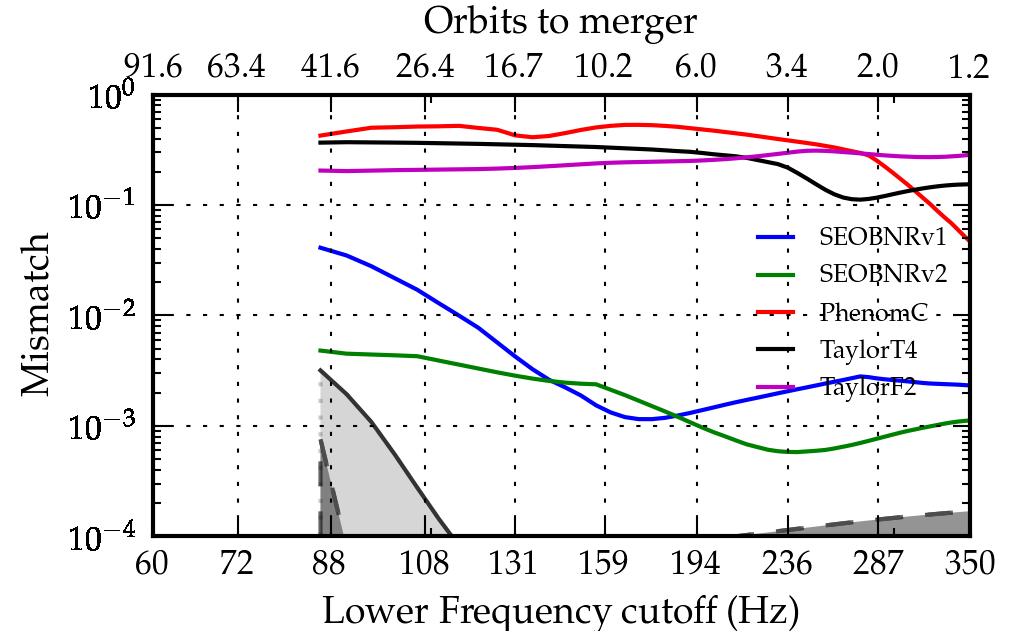}
\includegraphics[width=\columnwidth]{%
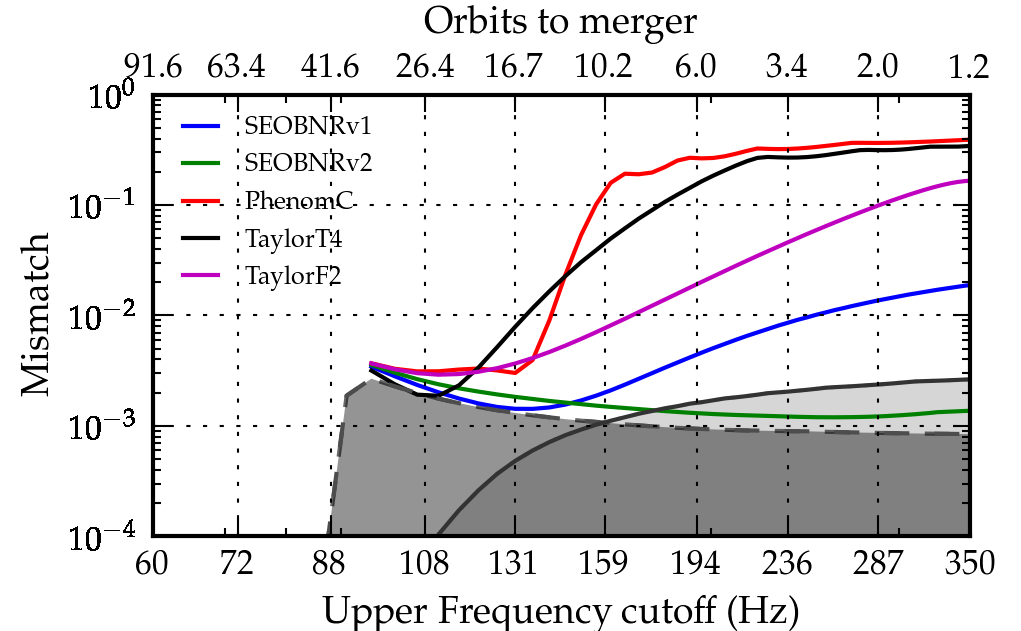}
\caption{These figures are similar to the two panels of
Fig.~\ref{fig:model_nrmismatch_freq_low_s06}, with the difference that the
system considered here has
$q=m_\mathrm{BH}/m_\mathrm{NS}=7M_\odot/1.4M_\odot =5$,
$\chi_\mathrm{BH} = -0.9$, and starts at GW frequency $\simeq 80$~Hz. This
corresponds to simulation ID~SXS:BBH:208 (see Table~\ref{table:simlist}).
Because of the width of the tapering window, we begin filtering at $86$~Hz. 
}
\label{fig:model_nrmismatch_freq_low_s-09}
\end{figure*}
\begin{figure*}
\centering
\includegraphics[width=0.8\columnwidth]{%
match_q7_s04_s0_60Hz_0042799_MismatchStartingFrequency_All.png}
\includegraphics[width=0.8\columnwidth]{%
match_q7_s04_s0_60Hz_0042799_MismatchEndingFrequency_All.png}\\
\includegraphics[width=0.8\columnwidth]{%
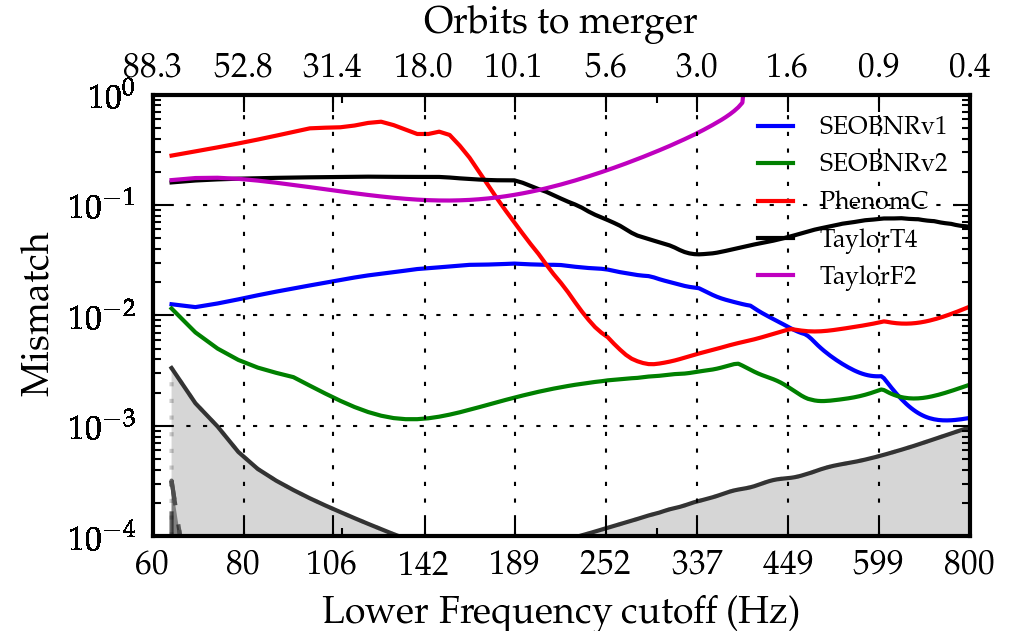}
\includegraphics[width=0.8\columnwidth]{%
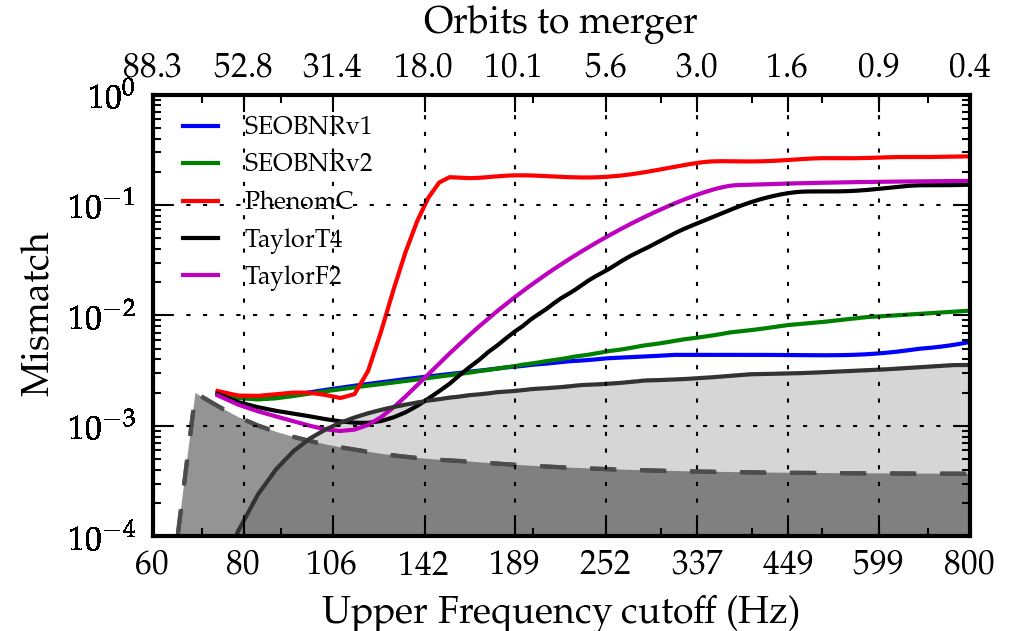}
\caption{The top panels in this figure are reproductions of the top two panels 
of Fig.~\ref{fig:model_nrmismatch_freq_low_s04}, and are shown here for direct 
comparison. 
The bottom two panels show identically computed quantities, with the only 
difference from the top two being that the NR waveform used corresponds to the 
second-highest numerical resolution instead of the highest. 
The case shown here has the highest mismatches between the highest and 
second-highest resolution NR waveforms, and therefore serves as a conservative 
example of the robustness of our results to NR errors.
}
\label{fig:model_nrmismatch_freq_low_s04_levs}
\end{figure*}
%

%
In this section, we use our long NR simulations, described in 
Sec.~\ref{s1:NRwaveforms}, to assess the accuracy of the different model 
waveforms over the inspiral and merger phases of NSBH binary coalescences.

\subsection{Mismatch Accumulation for Models}\label{s2:numrelmismatches}
In order to quantify GW model errors over different phases of binary
inspiral and merger, we compute matches between each model and NR
waveforms over cumulative frequency intervals. The results are shown
in
Fig.~\ref{fig:model_nrmismatch_freq_low_s06}-~\ref{fig:model_nrmismatch_freq_low_s-09}.
Our main results are as follows: (i) Of the two SEOBNR models,
SEOBNRv2 reproduces the late-inspiral and merger phases well for NSBH
binaries with
$-0.9\leq\chi_\mathrm{BH}\leq+0.6$. In contrast, the SEOBNRv1 model
has an erroneous phase evolution during the late-inspiral phase,
causing it to disagree both with the NR simulations and with
SEOBNRv2 (cf. Fig.~\ref{fig:seobv1v2_freq}) (ii) The pre-merger
phasing prescription of PhenomC does not reproduce NR waveforms
well, as is confirmed by Ref.~\cite{PhenomCWindowProblem}. (iii) Of the two
PN models we consider here, we found that TaylorT4 is more accurate
for aligned BH spins, while TaylorF2 is more accurate for 
anti-aligned BH spins.

In Figs.~\ref{fig:model_nrmismatch_freq_low_s06} 
and~\ref{fig:model_nrmismatch_freq_low_s04}, 
we show the mismatches between the TaylorT4, TaylorF2, PhenomC, 
SEOBNRv1, and SEOBNRv2 models and our aligned-spin simulations
with $q=m_\mathrm{BH}/m_\mathrm{NS}=9.8M_\odot/1.4M_\odot=7$, and 
$\chi_\mathrm{BH}=\{+0.6, +0.4\}$ (ID~SXS:BBH:202, SKS:BBH:203 and SXS:BBH:204,
c.f. table~\ref{table:simlist}), as a function of the
lower (left panels) and upper (right panels) frequency cutoffs in the 
match calculation. 
First, we observe that SEOBNRv2 shows good agreement with NR with
mismatches below $0.5\%$ over all $55-88$ orbits. We also find that
SEOBNRv1 agrees with NR over most of the inspiral orbits, but diverges
closer to merger, with mismatches reaching $10\%$ when considering the
last few orbits.  Therefore, we conclude that the disagreement
between the two SEOBNR models for positive aligned spins that was seen
in Fig.~\ref{fig:seobv1v2_freq} stems from the phasing errors of
SEOBNRv1.
Next, we find that the PhenomC model agrees well with the first few
tens and the last few orbits of the NR waveforms only. It accumulates 
\textit{significant} phase errors in a narrow frequency band around 
$130-150$~Hz, with mismatches rising above $10\%$.
This explains the pattern of mismatch accumulation we observed in 
the middle panel of Fig.~\ref{fig:seobv2phenomC_freq} and reaffirms our 
conclusion that the pre-merger phasing prescription of PhenomC needs 
to be revisited for NSBH parameters~\cite{PhenomCWindowProblem}.
We also find that both TaylorF2 and TaylorT4 (c.f. right panels of 
Fig.~\ref{fig:model_nrmismatch_freq_low_s06} 
and~\ref{fig:model_nrmismatch_freq_low_s04}) agree with NR well up to the last
$15$ or so pre-merger orbits. Closer to merger, their mismatches smoothly rise
to $10\%$, which is expected as the PN approximation degrades with increasing
binary velocity.
In addition, we find that TaylorT4 agrees with these NR waveforms to 
higher frequencies than TaylorF2, which is consistent with 
Fig.~\ref{fig:seobv2t4_faith}, where we show that TaylorT4 best agrees with
SEOBNRv2 for aligned, moderate BH spins.
Finally, we note that the bottom row of 
Fig.~\ref{fig:model_nrmismatch_freq_low_s04} shows the 
comparison of different approximants with a shorter NR simulation 
(ID~SXS:BBH:203, starting frequency $\sim 80$~Hz) with the same physical 
parameters as a longer simulation (ID~SXS:BBH:204, starting frequency
$\sim 60$~Hz). Therefore, the 
results for this simulation confirm those shown for the longer simulation 
(same figure, top row) for frequencies above $80$~Hz.

Next, we consider the case of NSBH binaries with anti-aligned BH spins. We 
show the mismatches between  approximants and NR waveforms
corresponding to $q=m_\mathrm{BH}/m_\mathrm{NS}=9.8M_\odot/1.4M_\odot =7$ 
and $\chi_\mathrm{BH} = \{-0.4,-0.6\}$ (ID~SXS:BBH:205-207), in Fig.~\ref{fig:model_nrmismatch_freq_low_s-04}
and Fig.~\ref{fig:model_nrmismatch_freq_low_s-06}, respectively.
We find that SEOBNRv2 shows the best agreement with NR, with 
mismatches below $0.2\%$ across all orbits. 
SEOBNRv1 \textit{monotonically} accumulates increasing mismatches
when we lower the lower frequency cutoff and appears to do the 
same over the later orbits when we increase the upper frequency cutoff, 
suggesting a phasing mismatch between the inspiral and merger portions 
of the waveform.
Therefore, we conclude that the disagreement between the two SEOBNR models 
for anti-aligned spins that was seen in Fig.~\ref{fig:seobv1v2_freq} stems 
from the phasing errors of SEOBNRv1.
Similar to the aligned-spin cases, the PhenomC model agrees well with NR
over the first few tens and the last couple of orbits, with mismatches below
$1\%$, but accumulates large mismatches ($10\%$) over a relatively narrow 
frequency range around $110$~Hz. All of these observations are
qualitatively similar to the aligned-spin cases.
Lastly, we find that TaylorF2 shows excellent agreement with our NR waveforms
over the first $60-65$ orbits, with mismatches below $1\%$. It gradually 
diverges during the last $5$ pre-merger orbits, with high mismatches, which 
is consistent with the right panel of Fig.~\ref{fig:seobv2phenomcf2_faith}.
TaylorT4, on the other hand, performs worse, with mismatches accumulating
earlier on and rising to $10\%$.
The bottom row of Fig.~\ref{fig:model_nrmismatch_freq_low_s-04} shows identical
comparisons with the shorter NR waveform ID~SXS:BBH:205 with the same physical 
parameters as ID~SXS:BBH:206. As in the aligned case, 
these panels agree with and provide a confirmation for 
the results shown in the top row of the figure, for frequencies above $80$~Hz.

Finally, we compare different approximants with NR simulation ID~SXS:BBH:208,
which has 
a smaller mass-ratio $q=5$ but the most extremal BH spin of all the cases we 
consider, with $\chi_\mathrm{BH}=-0.9$. Compared to the above two anti-aligned 
spin cases, we find that the SEOBNRv1 mismatches rise to $5\%$, indicating that 
the phasing errors of SEOBNRv1 get worse with BH spin magnitude. SEOBNRv2 still 
reproduces the NR waveform the best, while PhenomC and both PN approximants 
show similar patterns to the $q=7$ anti-aligned spin cases.

Overall, we conclude that the more recently calibrated 
SEOBNRv2~\cite{Taracchini:2013rva} model reproduces the late-inspiral and 
merger phases well for NSBH binaries with $-0.9\leq\chi_\mathrm{BH}\leq+0.6$.
This model was calibrated to $8$ non-spinning and $30$ non-precessing spinning
simulations with mass-ratios up to $q=8$, and most of the NSBH systems we 
consider here are within the calibration range of the model. Therefore, we 
conclude that SEOBNR performs well when interpolated within its calibration 
range.
%
%
We also conclude that PhenomC does not reproduce NR waveforms 
well, and its NR-calibrated portion needs to be 
revisited~\cite{PhenomCWindowProblem}. Of the two PN models we 
consider here, TaylorT4 and TaylorF2, we found that the former has better 
accuracy for aligned-spin binaries, while the latter is more accurate for the 
case of anti-aligned BH spins. These patterns, however, are likely 
coincidental.

Finally, note that in detection searches there is an additional degree of
freedom corresponding to the maximization of the SNR over the intrinsic 
waveform parameters. We investigate the suitability of the GW models 
considered here as detection templates in Sec.~\ref{s1:parameterbias}.

\subsection{Impact of numerical errors in NR simulations}\label{s2:nrerrorimpact}
Fig.~\ref{fig:ConvergenceTest} shows that some of the NR simulations (especially 
the ones with $\chi_\mathrm{BH}=\pm 0.4$) do not show explicit convergence with 
an increase in numerical grid resolution. Therefore, one might question the 
effect of the errors in the numerical waveforms that arise because of a finite grid 
resolution. In order to quantify this, in this section, we repeat the overlap 
calculations for the $\chi_\mathrm{BH}=+0.4$ configuration, using the NR 
waveform produced at the second-highest grid resolution. We 
choose this configuration because it exhibits the highest NR(highest 
resolution)-NR(next-to-highest resolution) mismatches, 
as shown by the solid line bounding the 
light grey region in Fig.~\ref{fig:model_nrmismatch_freq_low_s04}. 
%
We show the results in the bottom row of 
Fig.~\ref{fig:model_nrmismatch_freq_low_s04_levs}. The top row in the figure is 
a reproduction of the top row of Fig.~\ref{fig:model_nrmismatch_freq_low_s04}, 
which uses the NR waveform produced at the highest grid resolution. From the 
left two panels, we first notice that the PhenomC, TaylorT4 and TaylorF2 
mismatches are sufficiently large for the variations because of NR waveform 
differences to be inconsequent. 
Further, for both SEOBNRv1 and SEOBNRv2, we find 
a small change in the mismatches near the left and the right edge of the two 
left panels. These fluctuations are entirely consistent with the value of the 
NR-NR mismatch (shown by the solid grey line), 
when we apply the triangle inequality to the square roots of the SEOBNR-NR and 
NR-NR mismatches (see Sec.~III of Ref.~\cite{Kumar:2013gwa} for a brief 
discussion on manipulation of waveform mismatches arising from independent 
sources). The same is true when we compare the two right panels in 
Fig.~\ref{fig:model_nrmismatch_freq_low_s04_levs}. In this case, the changes 
visible in the SEOBNR-NR mismatches are below the upper bound set by adding the 
square roots of the SEOBNR-NR and NR-NR mismatches.

We therefore conclude that the fluctuations in the mismatches computed in the 
previous subsection are within the error bounds set by comparing NR waveforms at 
the highest and second-highest numerical grid resolutions. These bounds are 
shown explicitly in light grey shading bounded by solid lines in all panels of 
Fig.~\ref{fig:model_nrmismatch_freq_low_s06}-~\ref{fig:model_nrmismatch_freq_low_s-06}.
As these fluctuations remain below $0.3\%$ in all cases, our conclusions remain
robust to NR waveform errors.
%

\section{Effectualness and Parameter bias}\label{s1:parameterbias}
\begin{figure}
\centering
    \includegraphics[width=\columnwidth]{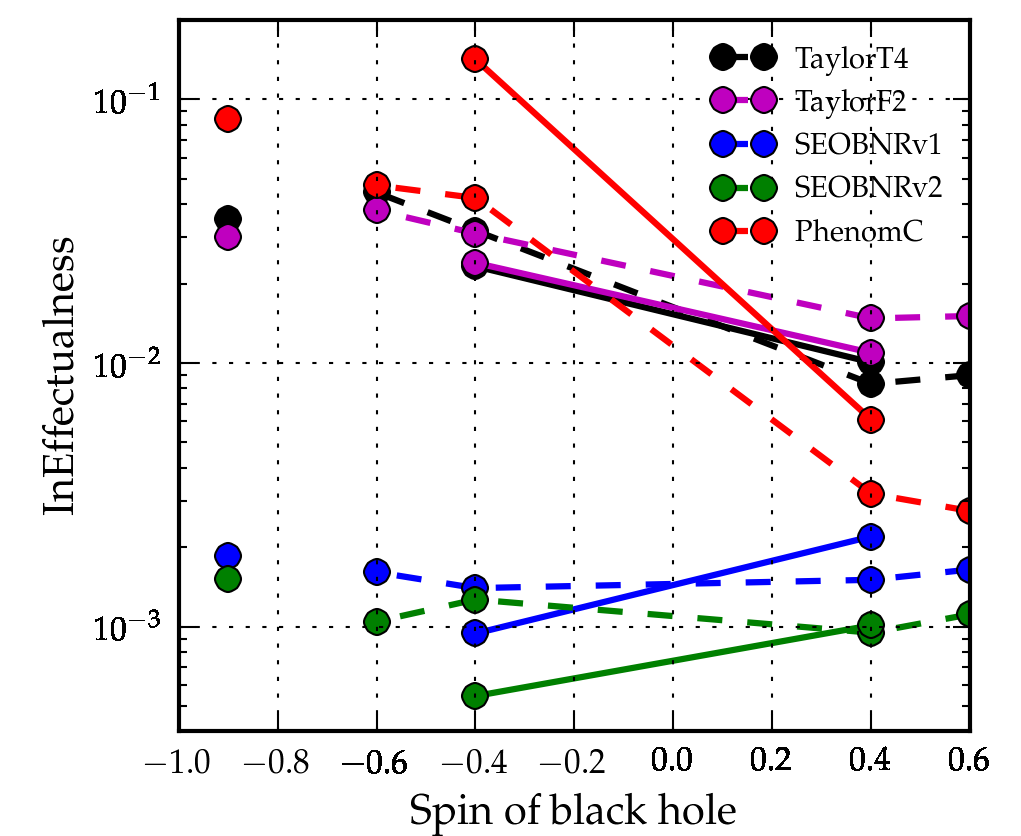}
\caption{In this figure, we show the ineffectualness of different waveform
approximants against our NR waveforms, as a function of black hole spin. 
Dashed lines join points corresponding to the cases where the NR waveform
starts at $\sim 80$~Hz (ID: 01, 02a, 03a, 04), while the solid lines 
correspond to the cases where the NR waveform starts at $\sim 60$~Hz 
(ID: 02b, 03b).
The isolated points correspond to the $q=5$ simulation (ID: 5).
We find that SEOBNRv2 consistently recovers the highest fraction 
($\geq 99.8\%$) of the optimal SNR when optimizing over intrinsic binary 
parameters. SEOBNRv1 is also fairly effectual.
The Taylor approximants recover $97-98\%$ of the SNR and noticeably more 
for the longer NR waveforms (for fixed binary parameters).
This is the expected trend, as longer waveforms extend to lower frequencies, 
where the PN approximation is better.
PhenomC is effectual against NR for $\chi_\mathrm{BH}\geq 0$, recovering
$\gtrsim 99\%$ of the SNR. For anti-aligned spins, however, its effectualness
was found to be low and decreasing with increasing NR waveform length. 
}
\label{fig:approx_nr_ineff}
\end{figure}
\begin{figure*}
\centering
    \includegraphics[width=0.66\columnwidth]{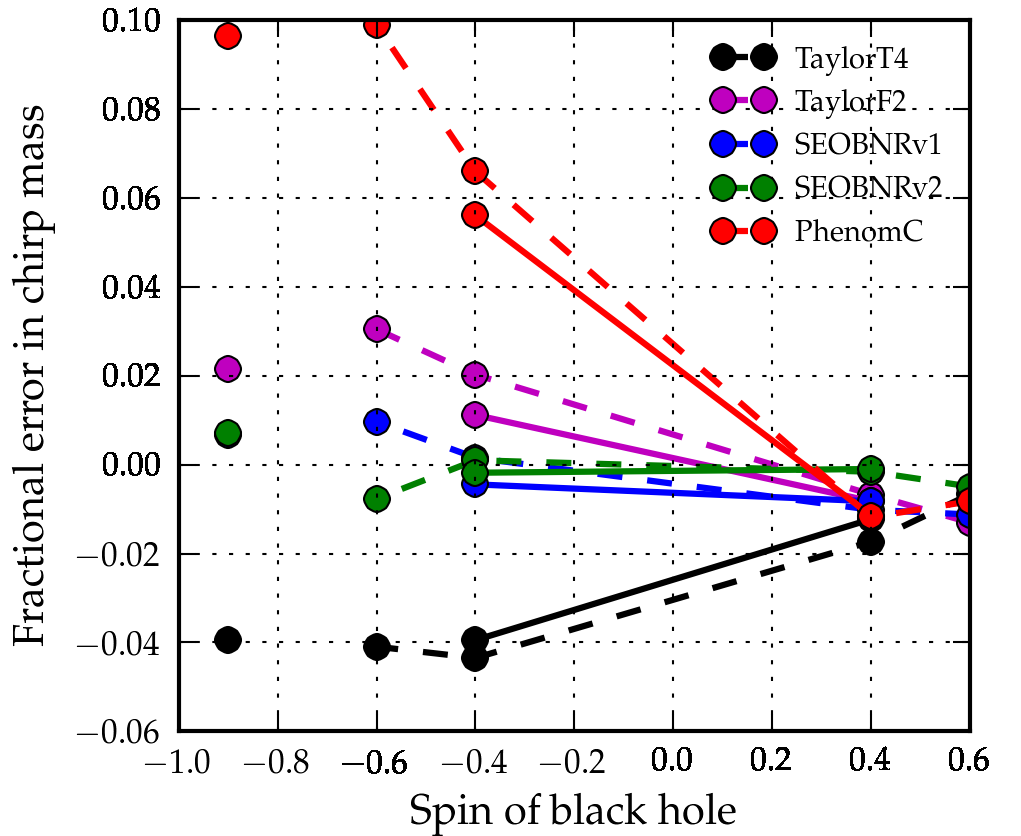}
    \includegraphics[width=0.66\columnwidth]{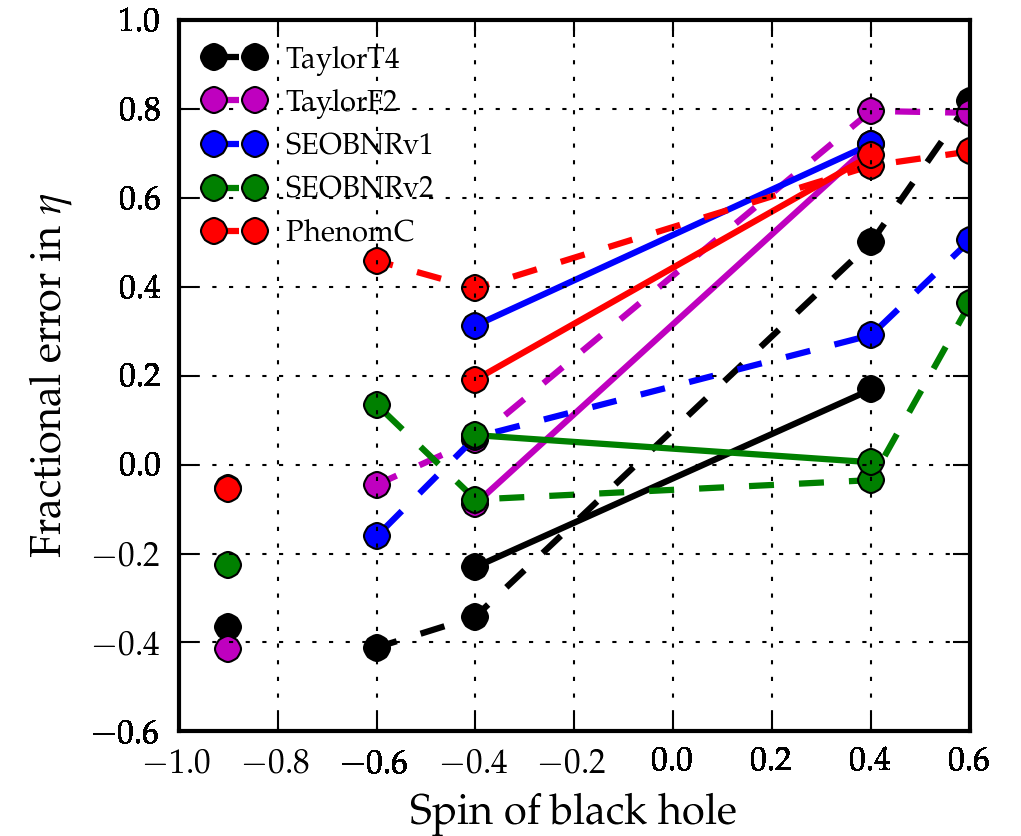}
    \includegraphics[width=0.66\columnwidth]{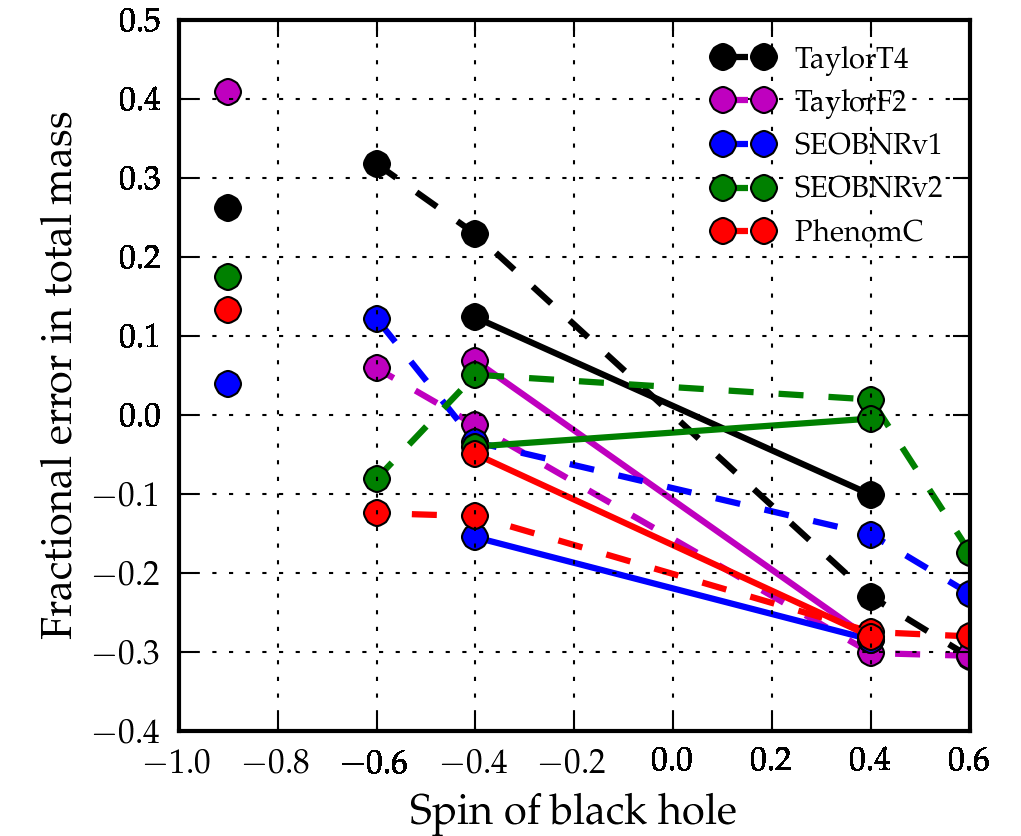}\\
    \includegraphics[width=0.66\columnwidth]{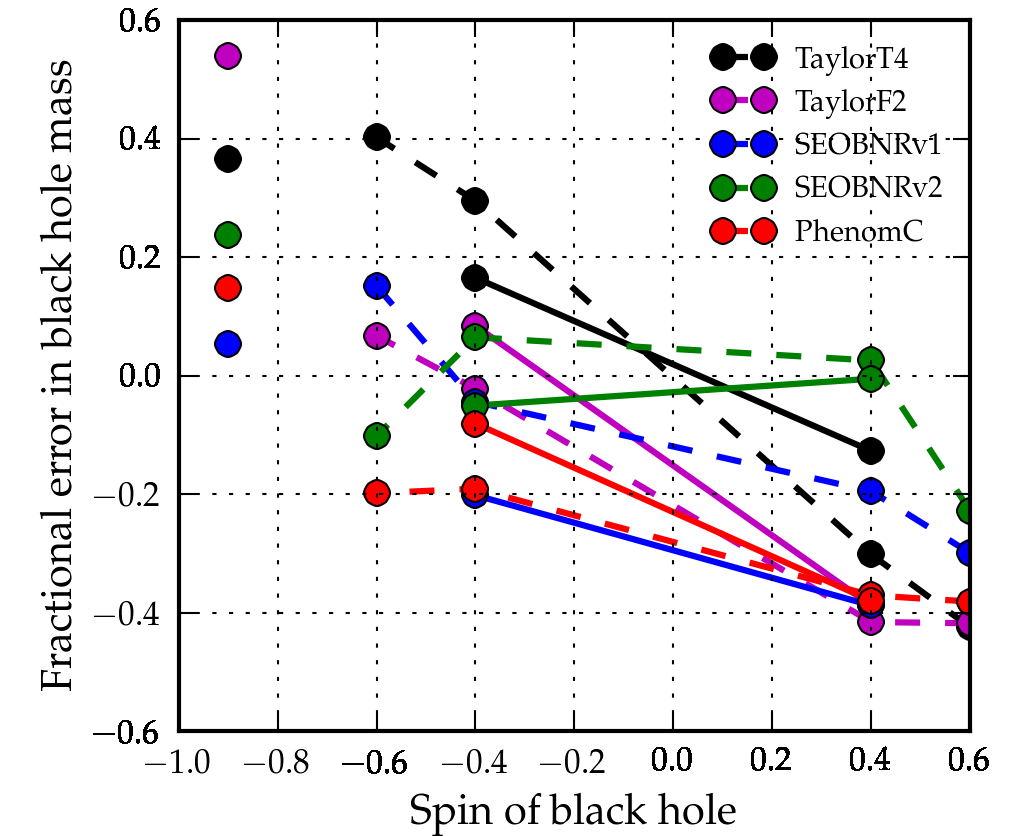}
    \includegraphics[width=0.66\columnwidth]{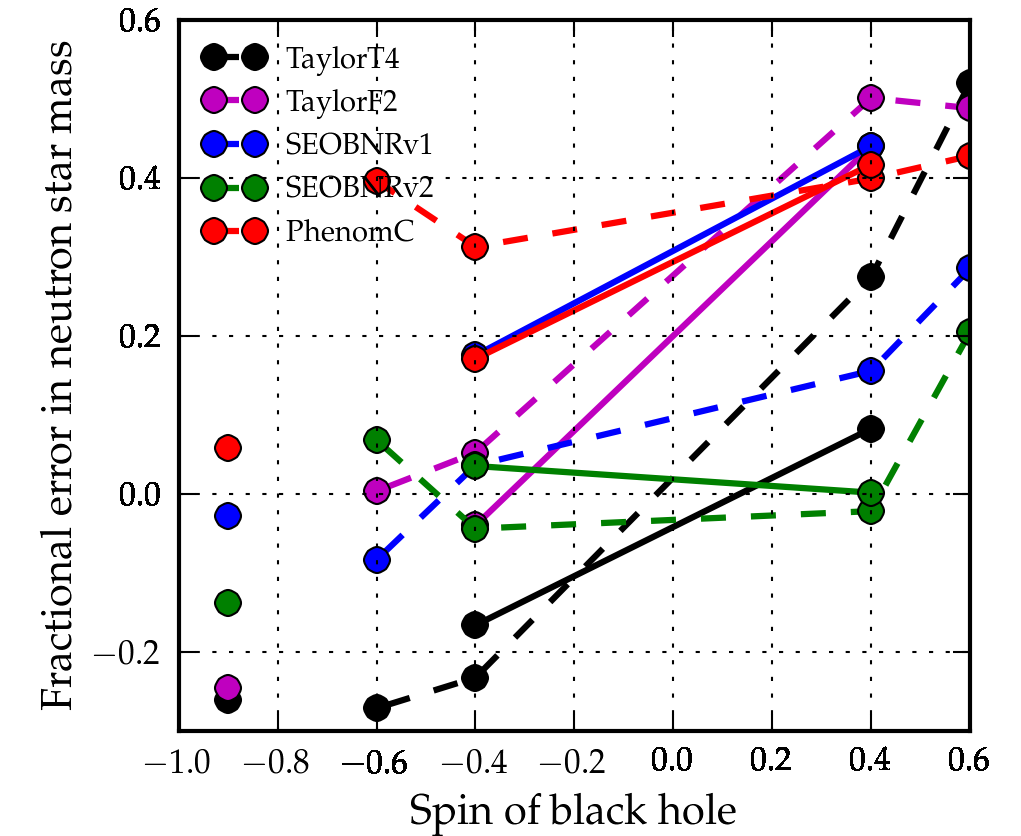}
    \includegraphics[width=0.66\columnwidth]{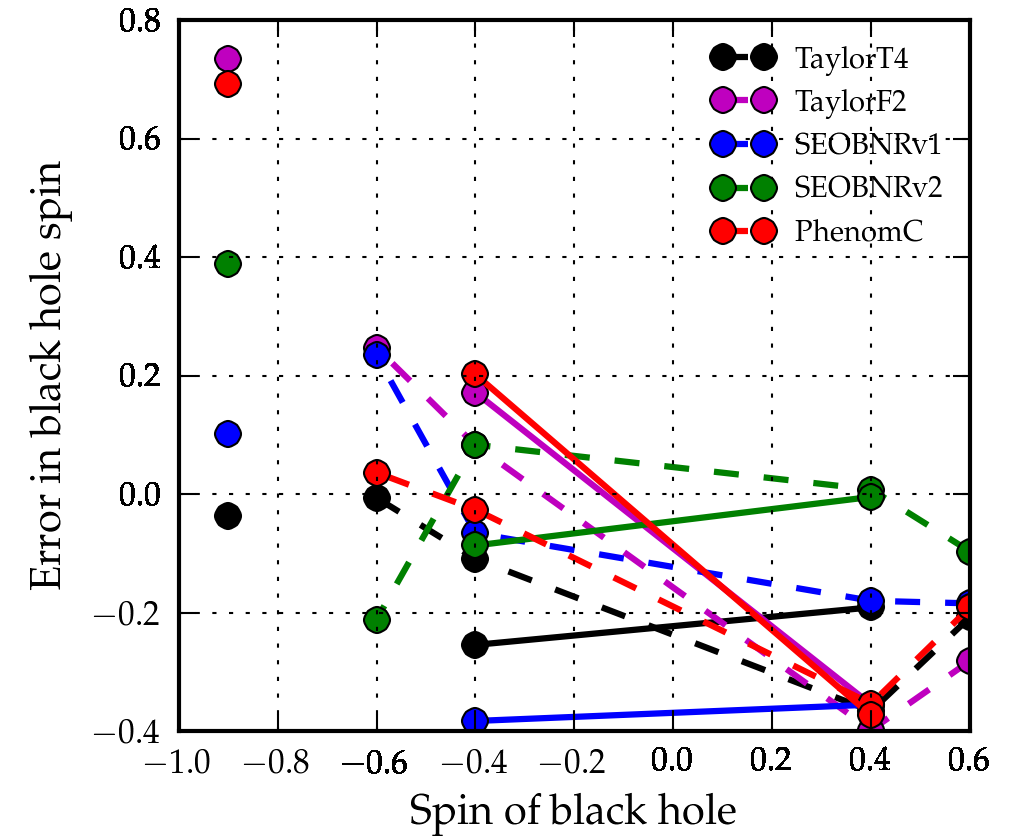}
\caption{These figures show the fractional error in the recovered maximum 
likelihood parameters as a function of black hole spin, except for black hole
spin for which the actual error value is shown. Each approximant is 
denoted with a unique color.
As in Fig.~\ref{fig:approx_nr_ineff}, dashed lines join points corresponding to 
NR waveforms that start at $\sim 80$~Hz (ID: 01, 02a, 03a, 04), while the 
solid lines correspond to the ones starting at $\sim 60$~Hz (ID: 02b, 03b). 
The isolated points correspond to the $q=5$ simulation (ID: 5).
While the left-to-right trends show us the effect of component spin, 
comparing dashed and solid lines show us the effect of including more inspiral
cycles.
}
\label{fig:param_bias_q7_sAll}
\end{figure*}
\begin{center}
\begin{table}[!htb]
\begin{tabular}{ c|c c c c c }
  \hline
  \multicolumn{6}{c}{SEOBNRv1} \\
  \hline
  Actual $S_{\mathrm{BH}}$ & $\mathcal{M}_c$ & $\eta$ & $m_1$ & $m_2$ & 
$S_{\mathrm{BH}}$ \\
  \hline
  0.6 & $-1.12\%$ & $50.45\%$ & $-29.9\%$ & $28.6\%$ & $-0.184$ \\
  0.4 (80Hz) & $-1.02\%$ & $29.1\%$ & $-19.5\%$ & $15.6\%$ & $-0.180$ \\
  0.4 (60Hz) & $-0.82\%$ & $72.1\%$ & $-38.8\%$ & $44.1\%$ & $-0.355$ \\
  -0.4 (80Hz) & $0.16\%$ & $6.19\%$ & $-4.40\%$ & $3.67\%$ & $-0.066$  \\
  -0.4 (60Hz) & $-0.44\%$ & $31.3\%$ & $-20.2\%$ & $17.6\%$ & $-0.382$ \\
  -0.6 & $0.97\%$ & $-16.1\%$ & $15.1\%$ & $-8.29\%$ & $0.234$ \\
  -0.9 $(q=5)$ & $0.66\%$ & $-5.2\%$ & $5.3\%$ & $-2.7\%$ & $0.102$ \\
  \hline \hline
  \multicolumn{6}{c}{SEOBNRv2} \\
  \hline
  Actual $S_{\mathrm{BH}}$ & $\mathcal{M}_c$ & $\eta$ & $m_1$ & $m_2$ & 
$S_{\mathrm{BH}}$ \\
  \hline
  0.6 & $-0.49\%$ & $36.3\%$ & $-22.8\%$ & $20.5\%$ & $-0.0966$  \\
  0.4 (80Hz) & $-0.16\%$ & $-3.48\%$ & $2.57\%$ & $-2.13\%$ & $0.00808$ \\
  0.4 (60Hz) & $-0.10\%$ & $0.54\%$ & $-0.51\%$ & $0.21\%$ & $0.00444$ \\
  -0.4 (80Hz) & $0.10\%$ & $-7.84\%$ & $6.48\%$ & $-4.35\%$ & $0.0836$ \\
  -0.4 (60Hz) & $-0.18\%$ & $6.64\%$ & $-5.03\%$ & $3.58\%$ & $-0.0864$ \\
  -0.6 & $-0.77\%$ & $13.5\%$ & $-10.2\%$ & $6.87\%$ & $0.234$ \\
   -0.9 $(q=5)$ & $0.7\%$ & $-22.7\%$ & $23.7\%$ & $-13.7\%$ & $0.387$ \\
\hline \hline
  \multicolumn{6}{c}{PhenomC} \\
  \hline
  Actual $S_{\mathrm{BH}}$ & $\mathcal{M}_c$ & $\eta$ & $m_1$ & $m_2$ & 
$S_{\mathrm{BH}}$  \\
  \hline
  0.6 & $-0.82\%$ & $70.5\%$ & $-38.1\%$ & $42.8\%$ & $-0.191$  \\
  0.4 (80Hz) & $-1.18\%$ & $67.3\%$ & $-37.0\%$ & $40.0\%$ & $-0.353$ \\
  0.4 (60Hz) & $-1.14\%$ & $69.6\%$ & $-38.0\%$ & $41.7\%$ & $-0.373$ \\
  -0.4 (80Hz) & $6.59\%$ & $39.7\%$ & $-19.1\%$ & $31.3\%$ & $-0.0259$ \\
  -0.4 (60Hz) & $5.61\%$ & $19.1\%$ & $-8.06\%$ & $17.1\%$ & $0.202$ \\
  -0.6 & $9.90\%$ & $45.9\%$ & $-19.8\%$ & $39.7\%$ & $0.0366$  \\
  -0.9 $(q=5)$ & $9.6\%$ & $-5.4\%$ & $14.8\%$ & $5.8\%$ & $0.693$ \\
  \hline \hline
\end{tabular}
\caption{In this table we list the fractional error in the recovered maximum 
likelihood parameters for different IMR approximants. For BH spin, we give
the actual difference between maximum likelihood and true parameter.
}
\label{tab:param_bias_IMR}
\end{table}
\end{center}
\begin{center}
\begin{table}[!htb]
\begin{tabular}{ c|c c c c c }
  \hline
  \multicolumn{6}{c}{TaylorT4} \\
  \hline
  Actual $S_{\mathrm{BH}}$ & $\mathcal{M}_c$ & $\eta$ & $m_1$ & $m_2$ & 
$S_{\mathrm{BH}}$ \\
  \hline
  0.6 & $-0.67\%$ & $81.9\%$ & $-42.4\%$ & $52.1\%$ & $-0.207$ \\
  0.4 (80Hz) & $-1.73\%$ & $50.0\%$ & $-30.1\%$ & $27.5\%$ & $-0.366$ \\
  0.4 (60Hz) & $-1.23\%$ & $16.92\%$ & $-12.7\%$ & $8.31\%$ & $-0.191$ \\
  -0.4 (80Hz) & $-4.35\%$ & $-34.2\%$ & $29.5\%$ & $-23.2\%$ & $-0.109$ \\
  -0.4 (60Hz) & $-3.95\%$ & $-23.0\%$ & $16.5\%$ & $-16.6\%$ & $-0.254$ \\
  -0.6 & $-4.09\%$ & $-41.1\%$ & $40.2\%$ & $-27.1\%$ & $-0.00606$ \\
  -0.9 $(q=5)$ & $-3.94\%$ & $-36.5\%$ & $36.6\%$ & $-26\%$ & $-0.0374$ \\
  \hline \hline
  \multicolumn{6}{c}{TaylorF2} \\
  \hline
  Actual $S_{\mathrm{BH}}$ & $\mathcal{M}_c$ & $\eta$ & $m_1$ & $m_2$ & 
$S_{\mathrm{BH}}$ \\
  \hline
  0.6 & $-1.30\%$ & $79.1$ & $-41.8\%$ & $48.9\%$ & $-0.281$ \\
  0.4 (80Hz) & $-0.67\%$ & $79.6\%$ & $-41.5\%$ & $50.1\%$ & $-0.398$ \\
  0.4 (60Hz) & $-0.82\%$ & $72.1\%$ & $-38.8\%$ & $44.1\%$ & $-0.355$ \\
  -0.4 (80Hz) & $2.02\%$ & $5.61\%$ & $-2.20\%$ & $5.27\%$ & $0.0828$ \\
  -0.4 (60Hz) & $1.12\%$ & $-8.82\%$ & $8.42\%$ & $-3.94\%$ & $0.172$ \\
  -0.6 & $3.04\%$ & $-4.53\%$ & $6.74\%$ & $0.40\%$ & $0.247$ \\
  -0.9 $(q=5)$ & $2.16\%$ & $-41.5\%$ & $54\%$ & $-24.6\%$ & $0.735$ \\
  \hline
\end{tabular}\label{tab:param_bias_I}
\caption{This table is identical to Table~\ref{tab:param_bias_IMR}, but
for inspiral-only approximants.
}
\label{tab:param_bias_I}
\end{table}
\end{center}
\begin{figure}
\centering
    \includegraphics[width=\columnwidth]{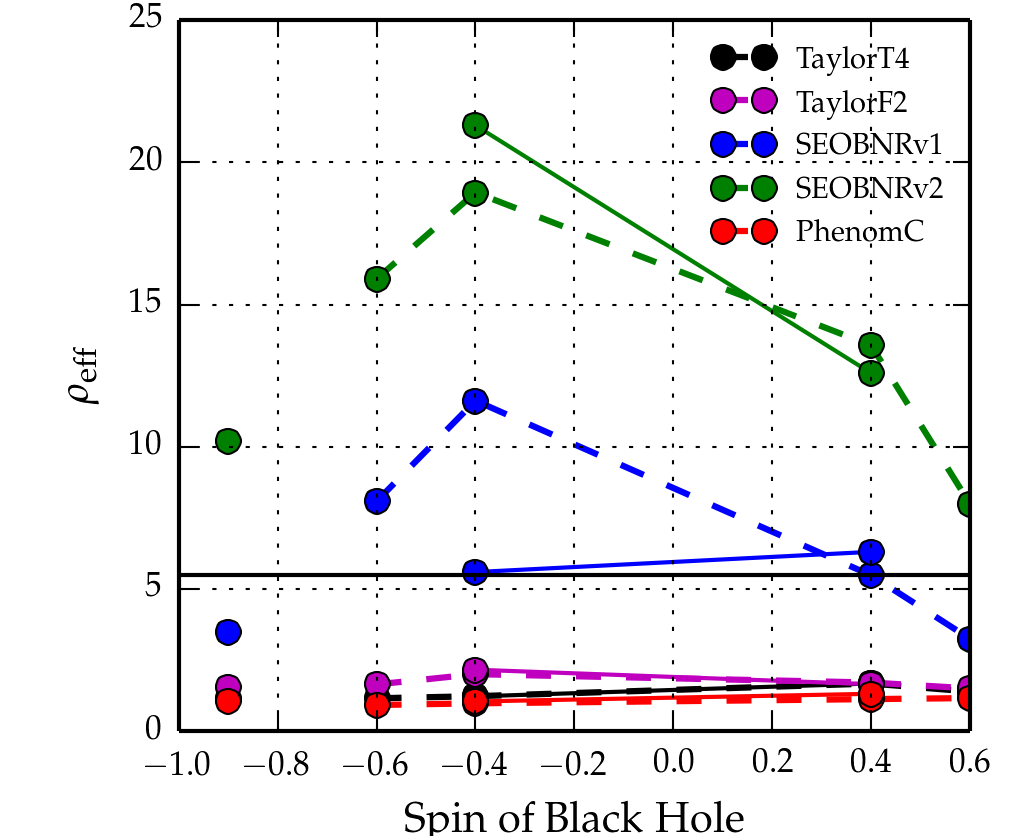}
\caption{In this figure, we show the lowest SNR value below which a modeled
waveform (using the respective approximant) with the same parameters as 
the true signal waveform will be indistinguishable from the 
true signal waveform. Here, the true signal waveform is represented 
by the corresponding NR waveform. We use
the criterion proposed in Ref.~\cite{WaveformAccuracy2008} for this 
calculation. As in Fig.~\ref{fig:approx_nr_ineff}, solid lines correspond to
the two longer simulations that start at $60$~Hz, dashed lines are
for the shorter simulations that start at $80$~Hz, and the isolated points
correspond to the q = 5 simulation (ID: 5). For all but SEOBNRv2 
(and marginally SEOBNRv1), this threshold is below the SNR cutoff ($=5.5$, 
shown by horizontal black line) employed by past 
LIGO-Virgo searches, and therefore their use would likely degrade the 
extraction of information from detector data.
}
\label{fig:snrEffective}
\end{figure}

While accurate parameter estimation requires that waveform models be
faithful to the true signal for given binary parameters, detection
searches allow for the additional degree of freedom of maximizing the
SNR over intrinsic binary parameters. With this freedom, intrinsic
waveform model errors can be compensated by slight shifts in the
physical parameters. As a result, a GW signal will be better matched
with a template waveform with slightly incorrect physical
parameters. In this section, we investigate the recovery of SNR using
different approximants and the associated biases in the recovered
maximum likelihood estimates for binary masses and spins.
We take all of our NR waveforms, produced at the highest numerical resolution, 
and compute their overlaps against 500,000 modeled waveforms with physical 
parameters in the vicinity of the true NR parameters. 
We compute waveform overlaps integrating from the starting frequency of the 
simulation in question, up to the Nyquist frequency corresponding to the 
waveform sampling rate.

In Fig.~\ref{fig:approx_nr_ineff} we show the fractional loss in SNR, or the 
ineffectualness of approximants, as a function of BH spin. The $q=7$ cases are 
connected with straight lines, while the single $q=5$ case is shown by a point.
Each approximant is shown with a different color. Solid and dashed lines join 
points corresponding to ID~$2a,3a$ and ID~$1,2b,3b,4$, respectively. 
Detection searches use banks of templates that correspond to a grid in the 
parameter space, and the SNR loss due to the discreteness of the grid will be 
\textit{in addition} to those shown in this figure. We find that both EOB 
models recover more than $99.5\%$ of the optimal SNR for both aligned and 
anti-aligned BH spins, despite different faithfulness. This is a good example 
of a shift in waveform parameters compensating for model phasing errors.
For aligned-spin cases, both PN approximants as well as PhenomC are effectual 
with SNR recovery above $98\%$. For all of the anti-aligned spin cases, however, 
we found relatively low SNR recovery using PN approximants, which is to be 
expected as anti-aligned spin binaries merge at lower frequencies than 
aligned-spin cases and therefore have relatively less signal power in the 
inspiral cycles above a given physical frequency (here $60-80$~Hz). PhenomC 
also shows significantly low effectualness for anti-aligned systems, recovering 
$\leq 96\%$ of the SNR for $q=7$ cases and $91.5\%$ for the 
$q=5, \chi_\mathrm{BH}=-0.9$ case. We therefore conclude 
that (a) both SEOBNR models are sufficiently accurate to model NSBH templates in 
aLIGO detection searches, and (b) PN approximants are also 
viable for use as filter templates for \textit{aligned}-spin NSBH systems.

In Fig.~\ref{fig:param_bias_q7_sAll} we show the fractional difference between 
the parameters that maximize the SNR recovery for each of the approximants, 
i.e. the maximum likelihood parameters, and the true physical parameters of the
system. Table~\ref{tab:param_bias_IMR} lists the
same for the IMR approximants, and Table~\ref{tab:param_bias_I} for the 
inspiral-only approximants. 
These differences quantify the systematic bias intrinsic 
to each approximant.
%
From the top left panel of Fig.~\ref{fig:param_bias_q7_sAll}, we observe that 
as the number of waveform cycles increases (monotonically with
BH spin), so does the accuracy of the recovered chirp mass. For aligned spins, 
all approximants but SEOBNRv2 converge at a systematic $-1\%$
bias in chirp mass recovery. SEOBNRv2 shows a smaller ($<0.8\%$) bias in chirp 
mass for all BH spins.
In the top center panel of Fig.~\ref{fig:param_bias_q7_sAll}, we show the bias 
in mass-ratio recovery. The spin-mass-ratio degeneracy that enters at the 
sub-leading order in PN phasing is manifest here, and for PN models, 
we find i) 
a larger bias in 
$\eta$ for aligned spins that increases with the spin magnitude and ii) 
a smaller (or negative) bias for anti-aligned spins. We see a similar trend
for the IMR approximants, with SEOBNRv2 showing the smallest systematic bias.
We also show biases in other binary mass combinations, total mass in the top 
right panel, BH mass in the bottom left, and NS mass in the bottom center. 
These show that none of the individual masses or their sum are nearly 
as accurately measured as the chirp mass of the binary. Additionally, the mass 
of the smaller component, here the NS, is slightly more biased than 
for the more massive component (here the BH).
Finally, the bottom right panel of the same figure shows the bias in recovered 
values of black hole spin. We observe a systematic underestimation of BH spin 
when it is aligned with the orbit, and smaller underestimation or 
overestimation when it is anti-aligned. Both of these patterns are exacerbated 
with BH spin magnitude. 

Overall, we found SEOBNRv2 to have the 
smallest systematic biases in parameter recovery. Therefore, we recommend its
use in aLIGO parameter estimation efforts focused on non-precessing NSBH 
binaries.

Having ascertained the systematic parameter biases that are applicable
in the limit of high SNR, we ask the question: how loud does an
incoming GW signal have to be before a modeled waveform with the same
parameters is distinguishable from the true, measured waveform? 
Ref.~\cite{WaveformAccuracy2008} proposed the criterion: $(\delta h|\delta h)<1$,
where $\delta h\equiv h^\mathrm{true} - h^\mathrm{approx}$,
which is sufficient for proving the indistinguishability of the modeled
waveform $h^\mathrm{approx}$ from the true signal $h^\mathrm{true}$.
We use it to calculate the 
effective SNR $\rho_\mathrm{eff}$ below which different approximants 
are indistinguishable from true (NR) waveforms, and show it in
Fig.~\ref{fig:snrEffective} as a function of black hole spin, for all
$q=7$ and $q=5$ cases. 
We immediately observe that for TaylorF2, TaylorT4, and PhenomC, the SNR 
threshold for distinguishability is below what is chosen as the single detector 
SNR lower 
cutoff in LIGO searches $(=5.5)$ and therefore using these approximants would 
likely degrade scientific measurements.
For SEOBNRv1, inclusion of more inspiral cycles for anti-aligned spin cases 
ID~$3,b$ cause a drop in $\rho_\mathrm{eff}$ from $12-6$. This is consistent 
with Fig.~\ref{fig:model_nrmismatch_freq_low_s-04} which shows that SEOBNRv1 
monotonically diverges from the reference NR waveform(s) when more of inspiral 
cycles are considered. For aligned-spin cases, we find that SEOBNRv1 is always 
distinguishable from a real signal with SNR above the lower cutoff for aLIGO 
searches.
SEOBNRv2 is indistinguishable from true waveforms to fairly high 
SNRs $\sim 15-20$ for anti-aligned spins, but this threshold lowers when we 
consider the longer, aligned-spin cases.

Lastly, we note that the SNR in consideration here is integrated from the 
starting frequency of the NR waveform in question, i.e. $\sim 60$~or~$80$~Hz, 
and corresponds to about $50-70\%$ of the total SNR accumulated over the entire
aLIGO frequency band starting from $15$~Hz.
These results are therefore likely to be pessimistic for
PN models, since these models do better at lower frequencies, 
where PN is more accurate.

\section{Conclusions}\label{s1:conclusions}

With the first observations of the Advanced LIGO and Virgo detectors 
imminent, rapid development of data analysis methods is underway within the
LIGO Scientific Collaboration and the Virgo Collaboration. 
Gravitational-wave astronomers are 
shaping matched-filtering-based targeted 
searches for neutron star - black hole binaries. As a 
step forward from most of earlier LIGO-Virgo searches 
(which used non-spinning waveforms as templates, 
e.g.~\cite{Colaboration:2011nz,Abadie:2010yb,Abbott:2009qj,Abbott:2009tt}), 
Advanced LIGO searches
plan to employ aligned-spin waveforms as templates. This is motivated
towards increasing the sensitivity of the searches for binaries with spinning
components. Recent work has shown that even if the component spins are not 
aligned to the orbital angular momentum and result in orbital precession, 
aligned-spin waveform templates would likely have significantly better 
sensitivity towards such systems than non-spinning waveform 
templates~\cite{Harry:2013tca}.

Recent progress in numerical relativity has allowed for faster and more 
accurate general-relativistic numerical simulations of inspiraling black 
holes, including the effect of component spin~\cite{Szilagyi:2014fna}. 
With these advances, more and longer simulations of compact binary motion
have become possible~\cite{Mroue:2013xna}. While the possibility of using
numerical relativity waveforms directly as search templates has been 
demonstrated~\cite{Kumar:2013gwa}, Bayesian parameter estimation efforts
will require the ability to generate template waveforms for arbitrary source
parameters. This is computationally prohibitive with the current NR 
technology, and therefore approximate waveform models are indispensible. 
Using strong-field information from numerical relativity, Effective-One-Body
(EOB) and phenomenological (Phenom) models have been developed and calibrated to 
accurately model the late-inspiral motion of compact binaries all the way 
through merger. The NR input here has 
been critical, since the post-Newtonian expressions that form the basis of all 
IMR models are perturbative expansions in the invariant velocity $v/c$, which 
become inaccurate in the strong field, rapid motion regime.
While the Phenom models are closed-form in frequency domain and therefore 
the least expensive to generate, reduced-order methods have been recently applied
to the EOB family to mitigate their computational cost~\cite{Purrer:2014fza}.

In this paper, we present $7$ new NR simulations, $6$
with $q=m_1/m_2=7$ and black hole spins $\chi_\mathrm{BH}=\{\pm 0.4, \pm 0.6\}$,
and $1$ with $q=5$ and $\chi_\mathrm{BH}=-0.9$. The spin of the smaller 
object (a black hole representing the neutron star) is held 
at $0$. For $\chi_\mathrm{BH}=\pm 0.4$, we perform two simulations each, one 
starting at a gravitational-wave frequency of $60$~Hz and the other starting at 
$80$~Hz (corresponding to a total mass of 
$1.4M_\odot+7\times1.4M_\odot=11.2M_\odot$). These span $36-88$ 
pre-merger orbits. For all other parameter values, our simulations start 
close to $80$~Hz when scaled to appropriate NSBH masses 
(cf.~Table~\ref{table:simlist}). 
Using these simulations, we study the accuracy of 
different waveform approximants and their effectualness as models for 
search and parameter estimation templates for the Advanced detector era.

Our investigation of 
the faithfulness of two inspiral-merger-ringdown models and 
two inspiral-only PN models
shows that both PN models become increasingly unfaithful with increasing BH 
spin magnitudes as well as with binary mass ratio, with overlaps falling 
below $50\%$.
This is consistent with a similar study~\cite{Nitz:2013mxa} and is 
indicative of the breakdown of the PN approximation with increasing binary 
velocity. We find that PhenomC disagrees in an even larger portion of the 
parameter space with SEOBNRv2, with overlaps above $0.9$ for near-equal mass 
binaries with spin magnitude below $0.3$. Somewhat surprisingly, we also find 
that the two SEOBNR models diverge \textit{significantly} for anti-aligned BH 
spins. 
Next, we investigate the GW frequency dependence of these model 
disagreements to disambiguate the portion of the binary coalescence process,  
that each model fails to capture accurately. As expected, the PN models 
describe the inspiral well but break down closer to merger. We found 
that PhenomC accumulates most of the mismatch against SEOBNRv2 close to the 
frequencies where it switches its phase prescription from one piece of a 
piece-wise continuous function to another. Lastly, for aligned spins, 
SEOBNRv1 accumulates phase differences close to merger, but for 
anti-aligned spins it agrees with SEOBNRv2 close to merger, with most of its 
mismatch being accumulated earlier on during the late inspiral. We present 
these results in detail in Sec.~\ref{s1:modelcomparison}.

When we study 
the mismatch accumulation of GW models as a function of frequency,  
we find that the PN models are faithful at the lower frequencies of our NR 
waveforms, and diverge close to merger. We find that PhenomC  
reproduces NR waveforms accurately during the last $2-5$ and first $20-30$ 
orbits, but accumulates significant mismatches over a span of $\sim 20$ 
orbits around $100$~Hz. 
For SEOBNRv1, we find that (a) for aligned-spin binaries, it slowly 
accumulates phase error over the last $\sim 5$ orbits with 
mismatches rising to $10\%$, but agrees well earlier on, but (b) for 
anti-aligned binaries, it agrees well with NR during the last $10-20$ orbits  
and diverges \textit{monotonically} when more inspiral orbits are included,  
with 
mismatches rising to $2\%$. In contrast, SEOBNRv2 reproduced all of our NR 
waveforms well, throughout the probed frequency range, with mismatches below 
$1\%$. We summarize these results in Sec.~\ref{s1:numrelcomparison}.

Further, we study the effectualness of these models as detection templates.
Detection searches allow for the additional freedom of maximizing the match
of template with signal over the intrinsic parameters of the templates, 
allowing for partial compensation of modeling errors. Using this freedom, we 
computed the mass-and-spin optimized overlaps between our set of numerical 
waveforms and those generated using different waveform approximants. For BH 
spins aligned with the orbital momentum, all models are effectual 
against our NR waveforms, with SNR recovery above $98\%$. For anti-aligned BH 
spins, we find that both PN models are less effectual, 
with SNR recovery dropping 
below $\lesssim 96.5\%$. This is expected, since our NR waveforms have fewer 
inspiral orbits for anti-aligned spins, even though they span similar frequency 
ranges. We find that PhenomC recovers the lowest SNRs (below $90-95\%$, 
depending on spin) for anti-aligned spins.
We find that both SEOBNR models recover more than $99.8\%$ of the optimal SNR 
when maximized over physical parameters. Therefore, we recommend using SEOBNR 
models to model non-precessing templates in aLIGO detection searches. 
We also note that the unpublished PhenomD model (an improvement over 
PhenomC) has shown promising results in terms of accurately capturing the
merger waveforms for high mass-ratio non-precessing
binaries~\cite{SebastianPhenomDNRDA}, and would therefore be another suitable
candidate for modeling search and parameter estimation templates in aLIGO era.

Finally, we also investigate the systematic bias in maximum likelihood 
recovered parameters that GW models will incur if used to 
model parameter estimation templates. We find that while the chirp mass is 
recovered increasingly and very accurately (within a percent) with 
increasing number of binary orbits in detector frequency band, the 
spin-mass-ratio degeneracy makes accurate determination of other parameters 
more difficult. For mass-ratio, we find that 
TaylorT4, TaylorF2, PhenomC, and SEOBNRv1 have a $20-50+\%$ bias, 
with SEOBNRv2 relatively the most faithful to NR with a $2-38\%$ bias. 
We also find that the models consistently underestimate BH spins 
for aligned-spin binaries, and overestimate (or slightly underestimate)
it for anti-aligned binaries. 
Overall, we find SEOBNRv2 to be the most faithful approximant to NR. 
As these biases are applicable in the high SNR limit, we also investigate the 
SNR limit below which different approximants are essentially indistinguishable 
from NR waveforms~\cite{WaveformAccuracy2008}. We find that for PN and 
Phenom models, this is never the case for signals with SNRs above $5$; but 
up to SNR $\rho_\mathrm{eff} \sim 10-22$ (depending
on spin), any further increase in the accuracy of SEOBNRv2 will not affect the 
extraction of scientific information from detector data. We describe 
these results in Sec.~\ref{s1:parameterbias}.
Therefore, for NSBH binaries with moderate spins, parameter estimation efforts
will benefit from using SEOBNRv2 templates in aLIGO era.
However, given the drop in accuracy of the SEOBNRv1 model outside its 
range of calibration, we also recommend further investigating SEOBNRv2  
at more extremal component spins, and we recommend trusting SEOBNRv2 
only within its calibration range.
%

We note that we ignore a good fraction ($35-45\%$, depending on BH spin) of 
the signal power by considering only frequencies above $60-80$~Hz. Therefore,
our results are accurate in the high-frequency limit and are likely to be
pessimistic for the PN-based inspiral-only models. We plan to extended the 
study to lower frequencies in future work.
%
We also note that there is tremendous ongoing effort to model the effect of
the tidal distortion of neutron stars during NS-BH 
mergers~\cite{Lackey:2013axa}. They are expected to be measurable with aLIGO
detectors~\cite{Read:2013zra}. However, matter effects affect binary phasing at 
5+PN order, while there are lower order unknown spin dependent terms in PN 
phasing whose lack of knowledge will have a larger impact on the detection 
problem~\cite{Ohme:2013nsa}. This motivates our choice to ignore 
neutron star matter effects in our numerical simulations, and instead treating
the neutron stars as low-mass black holes.
We have also ignored the effect of the possible tidal disruption of
NS in this study, and leave its detailed analysis to future work. 
However, since NS disruption affects NSBH waveforms only at very high
($\gtrsim 1.2$~kHz) frequencies~\cite{Foucart:2014nda} where aLIGO has 
significantly reduced sensitivity, ignoring it is unlikely to affect the 
accuracy of our analysis.
Finally, we note that we consider only the dominant $l=\pm m=2$ waveform 
multipoles in this study, since (i) other multipoles have much smaller (by 
orders of magnitude) contribution to the SNR, and (ii) none of the IMR
models considered here include sub-dominant modes.

\acknowledgments

We thank the Gravitational-Wave group at Syracuse
University for productive discussions. 
%
DAB and PK are grateful for hospitality of the TAPIR
group at the California Institute of Technology, where part of this
work was completed. PK acknowledges support through the Ontario
Early Research Award Program and the Canadian Institute for Advanced
Research. DAB and SB are supported by NSF awards 
PHY-1404395 and AST-1333142; NA and GL are supported by NSF award 
PHY-1307489 and by the Research Corporation for Science Advancement.
KB, MS, and BSz are supported by the 
Sherman Fairchild
Foundation and by NSF grants PHY-1440083 and AST-1333520 
at Caltech.
Simulations used in this work were performed with the \texttt{SpEC}
code~\cite{spec}.  Calculations were performed on the Zwicky cluster
at Caltech, which is supported by the Sherman Fairchild Foundation and
by NSF award PHY-0960291; on the NSF XSEDE network under grant
TG-PHY990007N; on the Syracuse University Gravitation and Relativity
cluster, which is supported by NSF awards PHY-1040231 and PHY-1104371
and Syracuse University ITS; 
on the Orca cluster suppoerted by NSF award PHY-1429873, the 
Research Corporation for Science Advancement, 
and by Cal State Fullerton;
and on the GPC supercomputer at the
SciNet HPC Consortium~\cite{scinet}. SciNet is funded by: the Canada
Foundation for Innovation under the auspices of Compute Canada; the
Government of Ontario; Ontario Research Fund--Research Excellence; and
the University of Toronto.  

\bibliographystyle{apsrev4-1}
\bibliography{paper}

\end{document}